\documentclass[aps,preprint,nofootinbib]{revtex4}
\usepackage{dcolumn}
\usepackage{bm}
\usepackage{amsthm}
\usepackage{color}

\usepackage{graphicx,epsfig}
\usepackage[english]{babel}
\usepackage{amssymb,amsfonts,amsmath}
\usepackage{verbatim}
\newcommand{\cA}{{\cal A}}

  \newcommand{\cH}{{\cal H}}
  \newcommand{\cJ}{{\cal J}}

\newcommand{\cO}{{\cal O}}  
  
  \newcommand{\cT}{{\cal T}}

\newcommand{\be}{\begin{equation}} \newcommand{\ee}{\end{equation}}
\newcommand{\bea}{\begin{eqnarray}} \newcommand{\eea}{\end{eqnarray}}
\newcommand{\beann}{\begin{eqnarray*}}  \newcommand{\eeann}{\end{eqnarray*}}
\newcommand{\bfig}{\begin{figure}} \newcommand{\efig}{\end{figure}}
\newcommand{\ba}{\begin{array}} \newcommand{\ea}{\end{array}}
\newcommand{\bcen}{\begin{center}} \newcommand{\ecen}{\end{center}}
\newcommand{\btab}{\begin{tabular}} \newcommand{\etab}{\end{tabular}}

\newcommand{\vev}[1]{\left\langle{#1}\right\rangle}

\newtheorem{Proposition}{Proposition}[section]

\newtheorem{Theorem}{Theorem}[section]
\newtheorem{Lemma}{Lemma}[section]
\newtheorem{Corrolary}{Corrolary}[section]

\newcommand{\bp}{\begin{Proposition}}   \newcommand{\ep}{\end{Proposition}}
\newcommand{\bt}{\begin{Theorem}}   \newcommand{\et}{\end{Theorem}}
\newcommand{\bl}{\begin{Lemma}}     \newcommand{\el}{\end{Lemma}}
\newcommand{\bc}{\begin{Corrolary}} \newcommand{\ec}{\end{Corrolary}}

\def\half{\frac{1}{2}}
\def\hzeta{{\hat\zeta}}
\def\tzeta{{\tilde\zeta}}

\begin{document}

\begin{flushright}
\end{flushright}

\title{Odd Parity Transport In Non-Abelian Superfluids From Symmetry Locking}

\author{Carlos Hoyos, Bom Soo Kim and Yaron Oz}
\affiliation{Raymond and Beverly Sackler School of
Physics and Astronomy, Tel-Aviv University, Tel-Aviv 69978, Israel\\
 E-mail: \tt{choyos,bskim,yaronoz@post.tau.ac.il}
}

\begin{abstract}
We consider relativistic non-Abelian superfluids, where the expectation value of the global symmetry currents relate
space and internal indices, thus creating  a ``locked" phase.  Locking a superfluid with $SU(2)$ internal symmetry in $2+1$ dimensions breaks parity
spontaneously, and introduces
parity-odd terms in the constitutive relations.
We show that there are
qualitatively different extensions of the rest frame locking to non-zero velocities.
We construct the resulting superfluid hydrodynamics up to the first derivative order.
Using an expansion close to the critical point, we estimate the ratio of the Hall
viscosity and the angular momentum density. Our general hydrodynamic results are compatible with the holographic $p$-wave
calculations in arXiv:1311.4882.

\end{abstract}

\maketitle


\tableofcontents

\newpage

\section{Introduction}
\label{sec:intro}

Breaking parity introduces new transport effects in fluid hydrodynamics. The parity breaking
can be explicit, spontaneous or as a consequence of quantum anomalies, each with its unique signature.
An interesting parity-odd transport is the dissipationless Hall viscosity $\eta_H$ in $2+1$ fluids \cite{Avron:1995,Avron:1997}. It enters in the constitutive relations of the fluid as a term in the stress tensor
\begin{equation}\label{Halltens}
T^{ij}_{\rm Hall} = \frac{\eta_H}{4}\left(\epsilon^{ik}\left(\partial^j v_k +\partial_k v^j)+\epsilon^{jk}(\partial^i v_k +\partial_k v^i\right)\right).
\end{equation}
Where $v_i$ is the (normal) velocity of the fluid. Latin indices $i,j,k$ refer to space components.
Its effect is to repel or attract nearby flows
due to a force perpendicular to the flow (for a recent review and references therein see \cite{Hoyos:2014pba}).
In most systems the value of the transport coefficient $\eta_H$ can be obtained from the two-point function of the stress tensor using the Kubo formula
\begin{equation}\label{kubo1}
\eta_H=\lim_{\omega\to 0}\frac{1}{4i\omega}\epsilon_{ik}\delta_{jl}\vev{T^{<ij>} T^{<kl>}}(\omega,\mathbf{k}=0).
\end{equation}
where the brackets $<ij>$ mean the traceless component and $\omega$, $\mathbf{k}$ are the frequency and momentum of the Fourier transformed correlators. This can be taken as the definition of Hall viscosity even in the cases where there is no hydrodynamic description.

Hall viscosity is present in diverse systems, particularly
in Quantum Hall states \cite{Avron:1995,Levay1995,Tokatly2007,Read:2008rn, Tokatly2009,Haldane2009,Read2011,Hoyos2012}, but also in other systems such as chiral and anyon superfluids in condensed matter \cite{Read:2008rn,Read2011,Hoyos2013}. In relativistic systems it has been found in holographic superfluids \cite{Saremi:2011ab,Son:2013xra} (see also \cite{Chen2011,Chen2012,Cai:2012mg}) and a related quantity, the `torsional Hall viscosity'\footnote{The distinction with the ordinary Hall viscosity is a bit subtle, the torsional Hall viscosity enters in the canonical energy-momentum tensor, which is not necessarily symmetric, while the ordinary Hall viscosity is defined for the symmetrized energy-momentum tensor. Despite their similar structure they are independent quantities.} has been found in `topological insulators'  in the presence of torsion \cite{Hughes:2011hv,Hughes:2012vg}.

It has been shown that a large number of systems exhibit an interesting relation between the Hall viscosity and the angular momentum density $\ell$, first derived
in \cite{Read:2008rn}, $\frac{\eta_H}{\ell} = \frac{1}{2}$.
The aim of this paper is to study parity-odd transport in superfluid hydrodynamics with spontaneously broken parity, and in particular the
general properties of the above relation between the Hall viscosity and the angular momentum density. Our general hydrodynamic results are compatible with the holographic $p$-wave
calculations in \cite{Son:2013xra}, where the relation between the Hall viscosity and the angular momentum density was argued to hold.

We will study the hydrodynamics of relativistic non-Abelian superfluids in a symmetry locked phase, that is
where the expectation value of the global symmetry currents relate
space and internal indices.  In particular, we will be locking a superfluid with an $SU(2)$ internal symmetry in $2+1$ dimensions, which breaks parity
spontaneously as a consequence of relating the $SU(2)$ structure constants to space structure.
We will define the locking when the normal component of the fluid is at rest, and we will show that there are various
qualitatively different extensions to non-zero velocities.
Locking will introduce
parity-odd terms in the hydrodynamic constitutive relations.  We will construct the resulting superfluid hydrodynamics up to the first derivative order.
Incidentally, we find that in the locked phases a term of the form \eqref{Halltens} appears in the stress tensor whose value is not determined by the Kubo formula \eqref{kubo1}, but by the two-point function between the stress tensor and the current
\begin{equation}
\widetilde{\eta}_H=\vev{\cO}+\frac{1}{4}\lim_{\omega\to 0}\delta_{ia}\delta_{jk}\vev{T^{<ij>} J^k_a}(\omega,\mathbf{k}=0).
\end{equation}
Where $\vev{J_a^i}=\vev{\cO}\delta_a^i$ is the expectation value of the $p$-wave condensate and $a=1,2$ are $SU(2)$ indices.
We have labelled this transport coefficient $\widetilde{\eta}_H$ to distinguish it from the usual definition \eqref{kubo1}. Using an expansion close to the critical point, we will estimate the ratio of the Hall
viscosity and the angular momentum density and find that
generically $\frac{\eta_H}{\ell}\sim  1$ for the standard Hall viscosity. On the other hand, we find that the ratio $\frac{\widetilde{\eta_H}}{\ell}$ depends on the type of locking.

The outline of the paper is as follows: In \S \ref{sec:hydro} we write down the constitutive relations for a general $SU(2)$ superfluid to first dissipative order. The locking can be seen as expanding around a particular value of the superfluid velocities. Since eventually we are interested in the physics close to the critical point, we keep terms only up to second order in the superfluid velocities. In \S \ref{sec:lock} we define the locking when the normal component of the fluid is at rest and study possible extensions to non-zero velocities. In \S \ref{sec:par} we identify the contributions to Hall viscosity and angular momentum in the locked phase and derive Kubo relations for them. In \S \ref{sec:holo}, we compare our predictions from the hydrodynamic analysis with the holographic $p$-wave model. We end by presenting our conclusions in \S \ref{sec:conc}. Some technical results are collected in the Appendices.

Here we provide three tables to guide the readers. The ideal constitutive relations and notations are collected in Table \ref{table:ideal}. Table \ref{table:projections} contains various projections of superfluid velocity discussed in \S \ref{sec:hydro}. Some definitions used in Kubo formulas in \S \ref{sec:par} are also listed in Table \ref{table:definitions}. $\mu,\nu = 0,1,2$ will denote space-time indices, $i,j=1,2$ space indices,  $A,B=1,2,3$ denote internal  $SU(2)$ indices, and
$a,b=1,2$ the internal indices without the direction $3$.

\begin{table}[ht]
\caption{Ideal order constitutive relations and notations}
\centering
\begin{tabular}{l c l}
\hline\hline
$T^{\mu\nu} =\varepsilon_n u^\mu u^\nu +p P^{\mu\nu}+f_{AB}\xi_A^\mu\xi_B^\nu$
& &  Stress-energy tensor \\
$J^\mu_A = q_A u^\mu -f_{AB} \xi^\mu_B$  & &  Internal symmetry currents \\
$\varepsilon_n+p =Ts+\mu_A q_A$
& \qquad\qquad\qquad  &  \\
$d\varepsilon_n = T ds+\mu_A dq_A +f_{AB}\xi^{\mu}_A d\xi_{B\,\mu}$  & &  Superfluid thermodynamics\\ [1ex]
\hline
$\varepsilon_n$ & & Energy density of the normal fluid component \\
$q_A$ & & Charge density of normal fluid component \\
$\xi_A^\mu$ & & Superfluid velocity \\
$\mu_A$ & & Chemical potential \\
$f_{AB}\! = \!f_0 \delta_{AB} +f_\mu \mu_A \mu_B$ & & \\
$q_s\delta^3_A =\! f_{AB}\mu_B =\!(f_0 +f_\mu \mu^2 ) \delta^3_A$ & & Charge density of the superfluid  component\\
$T$ & & Temperature \\
$s$ & & Entropy density  \\
$p$ & & Pressure \\
$\ell$ & & Angular momentum density\\[1ex]
\hline\hline
\end{tabular}
\label{table:ideal}
\end{table}

\begin{table}[ht]
\caption{Projections of the superfluid velocity}
\centering
\begin{tabular}{l c l}
\hline\hline
$P_{AB} = \delta_{AB} - \frac{\mu_A \mu_B}{\mu^2}$ & \hspace{1.5in}    & $C_{AB} = \epsilon_{ABC} \mu_C$    \\
$\zeta^\mu_A=P^\mu_{~\nu} \xi^\mu_A $ &   & $N^\mu =\frac{\zeta^\mu_A\mu_A}{\mu^2}$  \\
$\hzeta^\mu_A = \zeta^\mu_A - N^\mu \mu_A$ &  & $\zeta^{\mu\nu} = \hzeta^\mu_A \hzeta^\nu_A $\\
$\tzeta^\mu_A = \hzeta^\mu_B C_{AB} $ &   & $\tzeta^{\mu\nu} = \hzeta^\mu_A \tzeta^\nu_A $\\
$\zeta_{AB} = \hzeta^\alpha_A \hzeta_{\alpha B} $ &  &  \\
\hline\hline
\end{tabular}
\label{table:projections}
\end{table}

\begin{table}[ht]
\caption{Definitions and Kubo formulas}
\centering
\begin{tabular}{l c l}
\hline\hline
$\eta_H=\lim_{\omega\to 0}\frac{1}{4i\omega}\epsilon_{ik}\delta_{jl}\vev{T^{<ij>} T^{<kl>}}(\omega,\mathbf{k}=0)$& & Standard Hall viscosity \\
$\widetilde{\eta}_H=\vev{\cO}+\frac{1}{4}\lim_{\omega\to 0}\delta_{ia}\delta_{jk}\vev{T^{<ij>} J^k_a}(\omega,\mathbf{k}=0)$ & & Locking dependent Hall viscosity\\
$J_a^i=
\vev{\cO}\delta^i_a$ & & $p$-wave condensate \\
$\vev{\cO}\sim \sqrt{(T_c-T)/T_c}\sim \epsilon$ & & Order parameter  \\
$ \ell=2\mu C_3 \frac{\mu}{T} + 2\mu C_4 \langle \mathcal O \rangle$ & & Angular momentum density \\
$C_3= -\frac{i}{2\alpha_0}\lim_{\mathbf{k}\to 0}\epsilon_{ij}\frac{\partial }{\partial k_j}\vev{J^i_3 J^0_3}(\omega=0,\mathbf{k})$\\
$C_3Q+C_4q_n=\frac{1}{2}\delta_{ia}\vev{J^i_3 J^0_a}(\omega=0,\mathbf{k}=0)$ \\
$Q=\left(1+\frac{\mu B_\mu}{T B_T}\right)\frac{\langle \mathcal O \rangle}{f_0 T}$ \\
$\alpha_0=\left(1+\frac{q\mu }{T B_T}\right)\frac{1}{f_0 T}$ \\
$B_X =\frac{\partial p}{\partial X} + \frac{\langle \mathcal O \rangle^2}{f_0^2}\frac{\partial f_0}{\partial X}, \quad X=(T, \mu)$ \\
\hline\hline
\end{tabular}
\label{table:definitions}
\end{table}

\section{Non-Abelian superfluid hydrodynamics}
\label{sec:hydro}

In order to describe the hydrodynamic behaviour of a non-Abelian relativistic superfluid, we will make a generalization of the two-fluid model
\cite{Landau,Putterman,Bhattacharya:2011eea,Bhattacharya:2011tra,Neiman:2011mj} \footnote{Non-Abelian relativistic normal fluids were analysed
in \cite{Neiman:2010zi}.}.
The motion of the superfluid is determined by the conservation equations in the presence of a background metric $g_{\mu\nu}$ and gauge field $A_{A\,\mu}$,
\begin{align}
&\nabla_\mu T^{\mu\nu}=F^\nu_{A\ \mu}J^\mu_A,\\
&D_\mu J^\mu_A=0.
\end{align}
$\nabla_\mu$ is the covariant derivative with respect to the background metric and
\begin{equation}
D_\mu J^\mu_A= \nabla_\mu J^\mu_A+\epsilon_{ABC}A_{B\,\mu} J^\mu_C.
\end{equation}
The field strength is defined as usual
\begin{equation}
F_{A\,\mu\nu}=\partial_\mu A_{A\,\nu}-\partial_\nu A_{A\,\mu}+\epsilon_{ABC} A_{B\,\mu} A_{C\,\nu}.
\end{equation}
The constitutive relations of the energy-momentum tensor and the current are
\begin{align}
T^{\mu\nu} &=\varepsilon_n u^\mu u^\nu +p P^{\mu\nu}+f_{AB}\xi_A^\mu\xi_B^\nu+\pi^{\mu\nu},\\
J^\mu_A &= q_A u^\mu -f_{AB} \xi^\mu_B+\nu_A^\mu.
\end{align}
Here $u^\mu$ is the velocity of the normal component, $P^{\mu\nu}=\eta^{\mu\nu}+u^\mu u^\nu$ is the projector in the transverse direction, $q_A$ is the normal charge density and $\varepsilon_n$ the normal energy density. $p$ is the pressure and $\xi_A^\mu$ are the contributions to the currents due to the spontaneous breaking, usually identified with the superfluid velocity. In general we can expand the coefficients $f_{AB}$ as
\begin{equation}
f_{AB}=f_0\delta_{AB}+f_\mu \mu_A\mu_B.
\end{equation}
The terms $\pi^{\mu\nu}$ and $\nu_A^\mu$ contain  derivatives of the velocities and densities.

The thermodynamic equations are
\begin{align}
\varepsilon_n+p &=Ts+\mu_A q_A,\\
d\varepsilon_n &= T ds+\mu_A dq_A +f_{AB}\xi^{\mu}_A d\xi_{B\,\mu}.
\end{align}
We complement the hydrodynamic equations with the Josephson condition
\begin{equation}\label{joseph}
u^\mu\xi_{A\,\mu}=\mu_A+H_A,\\
\end{equation}
where $H_A$ depends on derivatives of the densities and superfluid velocities.

Using the equation
\begin{equation}
\nabla_\mu T^{\mu\nu}u_\nu+\mu_A D_\mu J^\mu_A=F^\nu_{A\,\ \mu} J^\mu u_\nu,
\end{equation}
One can show that there is a conserved entropy current when $\pi^{\mu\nu}=\nu^\mu_A=H_A=0$, provided that\footnote{One needs to use the first law in the following gauge-invariant form:
$$
\nabla_\mu\varepsilon_n=T\nabla_\mu s+\mu_A D_\mu q_A+f_{AB}\xi^\nu_A D_\mu \xi_{B\,\nu}.
$$}
\begin{equation}
f_{AB} u^\mu \xi^\nu_A(D_\mu \xi_{B\,\nu}-D_\nu\xi_{B\,\mu}-F_{B\,\mu\nu})=0.
\end{equation}
In the Abelian case this matches with the definition of the superfluid velocity as the covariant derivative of the Goldstone boson $\xi_\mu=-\partial_\mu\varphi+A_\mu$. In the non-Abelian case, there can be additional non-linear terms
\begin{equation}\label{vseq}
D_\mu \xi_{A\,\nu}-D_\nu\xi_{A\,\mu}+\lambda_\xi\epsilon_{ABC}\xi_{B\,\mu}\xi_{C\,\nu}=F_{A\,\mu\nu}.
\end{equation}
If we ignore gauge invariance, the gauge potential $A_{A\,\mu}$ that determines the field strength would have as many independent components as $\xi_{A\,\mu}$, so the number of independent equations would be sufficient to fix the superfluid velocities. However, because of gauge invariance, the number of independent equations is smaller. In the absence of external sources, if $\lambda_\xi=0$ there can be a gradient part $\xi^\mu_A=-\partial^\mu \varphi_A$ that is not fixed by the equations of motion. If $\lambda_\xi \neq 0$, then the part that is not fixed is of the form of a pure gauge $SU(2)$ potential
\begin{equation}
\xi_{A\,\mu} =\frac{i}{\lambda_\xi}(g^{-1}\partial_\mu g)_A, \ \ g\in SU(2).
\end{equation}
This justifies the addition of the Josephson condition \eqref{joseph} to the hydrodynamic equations.

When the dissipative terms are non-zero, the canonical entropy current is defined as
\begin{equation}
J_s^\mu= s u^\mu-\frac{1}{T}\pi^{\mu\nu}u_\nu-\frac{\mu_A}{T}\nu_A^\mu.
\end{equation}
The divergence of the entropy current obeys
\begin{equation}\label{jseq}
\nabla_\mu J_s^\mu = -\pi^{\mu\nu}\nabla_\mu\left(\frac{u_\nu}{T}\right)+\nu_A^\mu \left(\frac{E_{A\,\mu}}{T}-D_\mu\left(\frac{\mu_A}{T} \right) \right)+\frac{H_A}{T} D_\mu(f_{AB}\xi_B^\mu) ,
\end{equation}
where we defined the electric field as $E_A^\mu=F_A^{\mu \nu}u_\nu$. In order to impose the condition that the divergence of the entropy current is non-negative
$\nabla_\mu J_s^\mu\geq 0$ we should be able to write the rhs of \eqref{jseq} as a sum of squares. This implies that the dissipative terms can only depend on
\begin{equation}\label{bbs}
\sigma^{\mu\nu}, \ \ v^\mu_A=P^{\mu\alpha}D_\alpha\left(\frac{\mu_A}{T}\right)-\frac{E^\mu_A}{T}, \ \ s_A=D_\mu(f_{AB}\xi^\mu_B), \ \ \theta=\nabla_\mu u^\mu ,
\end{equation}
where the strain rate tensor is
\begin{equation}
\sigma^{\mu\nu}=P^{\mu\alpha}P^{\nu\beta}(\nabla_\alpha u_\beta+\nabla_\beta u_\alpha-P_{\alpha\beta}\nabla_\sigma u^\sigma).
\end{equation}
In principle it might be possible to modify the entropy current in such a way that there would be more allowed first order terms than the ones we present in \eqref{bbs}.\footnote{We would like to thank the referee for pointing this out to us.} However, as we discuss in Appendix \ref{modifent}, even if such terms were present they would not affect to the analysis of the Hall viscosity and the angular momentum density. We will then keep the discussion with the canonical entropy current bearing in mind that more general non-Abelian superfluid hydrodynamics might be possible (if this were the case our analysis could be understood as a subclass of theories where some transport coefficients are zero).

\subsection{Superfluid velocities in non-Abelian theories}

We presented above a consistent set of hydrodynamic equations and thermodynamic relations, that reduces to a familiar form in the Abelian case with $\xi^\mu$ being the superfluid velocity. In the non-Abelian case the properties of the superfluid should depend on the pattern of symmetry breaking. For a group $G$ broken to a subgroup $H$, the Goldstone bosons parametrize a coset $G/H$. Let $g\in G$ and $h\in H$ be group elements, such that the coset is determined by the equivalence $g\sim gh^{-1}$. We define the superfluid velocity as an element in the algebra\footnote{Under a local transformation $g\to g h^{-1}$ the gauge field transforms as $A_\mu \to h A_\mu h^{-1}-ih\partial_\mu h^{-1}$.}
\begin{equation}
\xi_\mu = ig^{-1}\partial_\mu g+A_\mu.
\end{equation}
In order to describe a coset we have to demand that the hydrodynamics currents are invariant under a global transformation $\xi^\mu\to h^{-1}\xi^\mu h$ with $h\in H$. This means adding additional constraints on the superfluid velocities. We will not pursue this direction here but in the following we will study the case where the group is completely broken.

Another new characteristic compared  to the Abelian case is that the symmetry can be broken if the currents acquire a non-zero expectation value. In the
non-Abelian  case the components of $\xi^\mu_A$ do not simply map to the gradient of the Goldstones, but they describe more generally the expectation value of the current. We will discuss  this in more detail  in the next sections. For now we will focus on finding an appropriate parametrization of the superfluid velocities.

There are two marked directions both in real and internal space. In real space the marked direction is determined by the velocity of the normal component $u^\mu$, while in the internal space it is determined by the chemical potential $\mu_A$. We will decompose the superfluid velocities in the directions
parallel and transverse to both.

The completely parallel direction is determined by the Josephson condition \eqref{joseph}. The dissipative term $H_A$ will not be important for us, since we will use the decomposition of the superfluid velocity in order to classify the first order terms. Then, at the ideal order we have
\begin{equation}
\xi^\mu_A=-\mu_A u^\mu+\zeta^\mu_A, \ \ u_\mu\zeta^\mu_A=0.
\end{equation}
We further decompose $\zeta^\mu_A$ in the parallel and transverse directions to $\mu_A$:
\begin{equation}
\zeta^\mu_A=N^\mu \mu_A+\hzeta^\mu_A,
\end{equation}
where the Abelian component of the superfluid velocity is
\begin{equation}
N^\mu =\frac{\zeta^\mu_A\mu_A}{\mu^2},
\end{equation}
and $\hzeta^\mu_A\mu_A=0$. Note that for $SU(2)$ $\hzeta^\mu_A$ in $2+1$ dimensions has four independent components before using the equations of motion. In order to find a suitable parametrization we first define the `spatial velocity' vectors
\begin{equation}
v=(u_1 ,\ u_2), \ \ m=(\mu_1, \ \mu_2).
\end{equation}
Then, the transverse components can be written in matrix form as (the first column corresponds to $A=3$ and the first row to $\mu=0$):
\begin{equation}
\hzeta^\mu_A =\left(
\begin{array}{cc}
\zeta_\lambda (v^T \bar{\sigma}^\lambda m) & - \mu_3 \zeta_\lambda (v^T\bar{\sigma}^\lambda) \\
-u_0 \zeta_\lambda (\bar{\sigma}^\lambda m) & \mu_3 u_0\zeta_\lambda \bar{\sigma}^\lambda
\end{array}
\right),
\end{equation}
where $\lambda=0,1,2,3$ and the $\bar{\sigma}^\lambda$ matrices are defined as the identity and the Pauli matrices:
\begin{equation}
\bar{\sigma}= \{\mathbf{1},\sigma^1,i\sigma^2,\sigma^3\}.
\end{equation}
The four independent components of $\hzeta^\mu_A$ are parametrized by  $\zeta_\lambda$.

\subsection{Dissipative terms}

We are interested in the behaviour of transport coefficients close to the critical point between the normal and the superconducting phase.\footnote{We assume here that there is no first order phase transition.} This implies that $\zeta^\mu_A$ should be small, either because a large superfluid velocity will destroy the superconducting phase or because $\zeta^\mu_a$ acts as an order parameter and it should vanish as the critical point is approached.

We will perform an expansion for small $\zeta^\mu_A \sim \epsilon\ll 1$, this means that both $N^\mu$ and $\hzeta^\mu_A$ are small $\sim \epsilon$. Within this expansion we will construct all possible terms to $O(\epsilon^2)$ that lead to a consistent hydrodynamic theory. Note, that the transport coefficient themselves can also be expanded in the scalars $N^2$ and $\hzeta^\alpha_A\hzeta_{\alpha\, A}$.

We define the even and odd projectors in the directions transverse to the chemical potential
\begin{equation}
 P_{AB}=\delta_{AB}-\frac{\mu_A \mu_B}{\mu^2}, \ \ C_{AB}=\epsilon_{ABC}\mu_C.
\end{equation}
To order $O(\epsilon)$ we can use the following two-index combinations of the transverse superfluid velocity
\begin{equation}
\hzeta^\mu_A, \ \ \tzeta^\mu_A=C_{AB}\hzeta^\mu_B.
\end{equation}
To order $O(\epsilon^2)$ we can use the following independent combinations
\begin{equation}
\zeta^{\mu\nu}=\hzeta^\mu_A \hzeta^\nu_A, \ \ \tzeta^{\mu\nu}=\hzeta^\mu_A\tzeta^\nu_A, \ \ \zeta_{AB}=\hzeta^\alpha_A\hzeta_{\alpha \,B}.
\end{equation}
Note,  that $\zeta^{\mu\nu}=\zeta^{\nu\mu}$ and $\tzeta^{\mu\nu}=-\tzeta^{\nu\mu}$.

We will decompose the dissipative terms using the Abelian component of the velocity and the chemical potential
\begin{equation}\label{dissipterms}
\begin{split}
\pi^{\mu\nu} &=N^\mu N^\nu \Sigma_1+P^{\mu\nu}\Sigma_2+N^{(\mu}V_1^{\nu)}+\tau^{(\mu\nu)}+\zeta^{\mu\nu}\Sigma_T,\\
\nu^\mu_A &= N^\mu\mu_A \Sigma_3+N^\mu V_{1\, A}+V_2^\mu \mu_A+\hzeta^\mu_A \Sigma_{V,1}+\tzeta^\mu_A \Sigma_{V,2}+\tau^\mu_A,\\
H_A &=\mu_A\Sigma_4+V_{2\, A} ,
\end{split}
\end{equation}
where $V_A\mu_A=\tau^\mu_A\mu_A=0$. All the possible first order terms to $O(\epsilon^2)$ can be found in the Appendix \ref{sec:diss1}.

\section{Symmetry Locked phases}
\label{sec:lock}

One of our  goals is to understand the origin and how general is the relation between the Hall viscosity and angular momentum density found in \cite{Son:2013xra} for the holographic $p$-wave model. In this model there is a nonzero chemical potential and charge density that we can choose to be $\mu_3\neq0$, $q_3\neq0$. Lorentz and $SU(2)$ symmetries are then reduced to spatial rotations $SO(2)_S$ and the $U(1)_3$ subgroup that leaves the chemical potential invariant. The parity breaking terms appear in a broken phase, where the currents acquire and expectation value $\vev{J_a^i}\propto \delta_a^i $ in such a way that space and flavour indices are related. We dub this as the `locked' phase by analogy with the color-flavour locking phase of QCD \cite{Alford:2007xm}.
In this phase the remaining symmetries are spontaneously broken to a diagonal $U(1)$:
\begin{equation}
SO(2)_S\times U(1)_3\to U(1)_D.
\end{equation}
Therefore, we expect this theory to have a single Goldstone mode. The origin of parity breaking is easy to understand, the $SU(2)$ structure constants are epsilon tensors that break internal `parity' transformations. After the locking, this breaking is transferred to the spatial directions as well.
 We can consider a transformation acting on the components of an object with one internal index $V_a$ as $V_1\leftrightarrow V_2$. The theory has also initially parity symmetry $x_1\leftrightarrow x_2$.
When the locking is made, the components of the non-Abelian current become $J^i_a\sim \delta^i_a$, which is invariant only under a combination of the internal and parity transformations. However, the internal transformation is not a symmetry because there are terms with epsilon tensors $\epsilon^{ABC}$ that change sign. Then, the would-be parity symmetry allowed by the locking that is a combination  of space and internal symmetries is broken by the epsilon terms. This means that there is no additional $Z_2$ symmetry in the superfluid phase.

When the normal fluid is at rest $u^\mu=(1,0)$, we can describe the locked phase in the hydrodynamic regime by setting
\begin{equation}\label{lock0}
\mu_A=\mu \delta_A^3,\ \ q_A=q_n\delta_A^3,\ \ N^\mu=0,\ \  \zeta_\lambda=(\zeta_s,0,0,0).
\end{equation}
Then,
\begin{equation}
\hzeta^\mu_A = -\mu \zeta_s \left(
\begin{array}{cc}
0 & 0 \\
0 &  \delta^i_a
\end{array}
\right).
\end{equation}
Here $\zeta_s$ is proportional to the $p$-wave condensate, more precisely
\begin{equation}\label{vev}
J_a^i= f_0\mu \zeta_s\delta^i_a=\vev{\cO}\delta^i_a.
\end{equation}
If we demand that $\zeta_s$ is constant in the absence of sources, this means that in the equation of motion for $\xi^\mu_A$ \eqref{vseq} the coefficient of the non-linear term should vanish $\lambda_\xi=0$.

There are several possible extensions to non-zero velocities of the normal component that lead to qualitatively different results. We will distinguish between locking in the lab frame and locking in the rest frame of the fluid. We present them here and discuss in the next section how the Hall viscosity is affected by the locking.

\subsection{Locking in the lab frame}

We fix the locking to be \eqref{lock0} even at non-zero velocities. The transverse components of the superfluid velocity are in this case:
\begin{equation}
\hzeta^\mu_A = \mu \zeta_s \left(
\begin{array}{cc}
0 & - u_a \\
0 & u_0 \delta^i_a
\end{array}
\right),
\end{equation}
We can also write it as $\hzeta^\mu_A=\mu\zeta_s\hat P^\mu_A$.
At the ideal order the currents are
\begin{equation}
J^\mu_A = (q_n +q_s)\delta_A^3 u^\mu -\vev{\cO} \hat P^\mu_A ,
\end{equation}
where we have used \eqref{vev}, and we identified the superfluid charge density as
\begin{equation}\label{qs}
q_s=\mu(f_0 +f_\mu\mu^2).
\end{equation}

From \eqref{dissipterms}, we are left with the following dissipative terms
\begin{equation}
\begin{split}
\pi^{\mu\nu} &=P^{\mu\nu}\Sigma_2+\tau^{(\mu\nu)}+\zeta^{\mu\nu}\Sigma_T,\\
\nu^\mu_A &= \mu \delta_A^3 V_2^\mu+\hzeta^\mu_A\Sigma_{V,1}+\tzeta^\mu_A\Sigma_{V,2}+\tau^\mu_A,\\
H_A &=\mu \delta^3_A\Sigma_4+V_{2\, A},
\end{split}
\end{equation}
where $V_3=\tau^\mu_3 =0$.

After the locking, the basic building blocks allowed by the entropy equation \eqref{bbs} become (in the absence of external sources)
\begin{equation}\label{bbslock1}
\sigma^{\mu\nu}, \ \ v^\mu_A=P^{\mu\alpha}\partial_\alpha\left(\frac{\mu}{T}\right)\delta_A^3, \ \ s_A=\partial_\mu(-q_s u^\mu \delta_A^3+\vev{\cal O}\hat P^\mu_A), \ \ \theta=\partial_\mu u^\mu.
\end{equation}
We wrote explicitly all the first order terms that survive the locking to $O(\epsilon^2)$ in the Appendix \ref{sec:diss2}.

\subsection{Locking in the rest frame}

By `locking in the rest frame' we mean that the normal component of the fluid and the chemical potential point in the same direction (taking `time' to be the third direction). This can be achieved by setting $\mu_A=-\mu u_A/\sqrt{(u_0)^2+u_i^2}\equiv -\mu u_A/u$ ($u_3=u_0$), $N^\mu=0$ and $\zeta_\lambda=(\zeta_s,0,0,0)$. Note that the normalization of the chemical potential is necessary in order to keep the condition $\mu_A\mu_A=\mu^2$.

Then,
\begin{equation}
\hzeta^\mu_A =-\frac{\mu \zeta_s}{u}\left(
\begin{array}{cc}
u_k^2 & - u_0 u_a \\
-u_0 u^i & u_0^2\delta^i_a
\end{array}
\right).
\end{equation}
We can also write it as $\hzeta^\mu_A=-\frac{\mu\zeta_s}{u}\hat P^{\mu A}=-\frac{\mu\zeta_s}{u}(P^{\mu A}+\tilde P^{\mu A})$, where
\begin{equation}
\tilde P^{\mu A}=\left(
\begin{array}{cc}
0 & 0 \\
0 & u_k^2\delta^i_a-u^i u_a
\end{array}
\right).
\end{equation}
At the  ideal order the currents are
\begin{equation}
J^\mu_A = -(q_n +q_s)u^\mu \frac{u_A}{u}+\frac{\vev{\cO}}{u} \hat P^{\mu A},
\end{equation}
where we used \eqref{vev} and \eqref{qs}.

For this type of locking the dissipative terms \eqref{dissipterms} are
\begin{equation}
\begin{split}
\pi^{\mu\nu} &=P^{\mu\nu}\Sigma_2+\tau^{(\mu\nu)}+\zeta^{\mu\nu}\Sigma_T,\\
\nu^\mu_A &= -\mu \frac{u_A}{u} V_2^\mu+\hzeta^\mu_A\Sigma_{V,1}+\tzeta^\mu_A \Sigma_{V,1}+\tau^\mu_A,\\
H_A &=-\mu \frac{u_A}{u}\Sigma_4+V_{2\, A},
\end{split}
\end{equation}
where $V_A u_A=\tau^\mu_A u_A=0$. The basic building blocks allowed by the entropy equation \eqref{bbs} become
\begin{equation}\label{bbslock2}
\sigma^{\mu\nu}, \ \ v^\mu_A=-P^{\mu\alpha}\partial_\alpha\left(\frac{\mu}{T u}\right)u_A-\frac{\mu}{T u}P^{\mu\alpha}\partial_\alpha u_A, \ \ s_A=\partial_\mu\left(q_s u^\mu \frac{u_A}{u}-\frac{\vev{\cal O}}{u}\hat P^{\mu A}\right), \ \ \theta=\partial_\mu u^\mu.
\end{equation}
The allowed first order terms to $O(\epsilon^2)$ for this kind of locking can be found in the Appendix \ref{sec:diss3}.

\section{Parity breaking effects}
\label{sec:par}

As we discussed, the locking will introduce parity breaking terms in the constitutive relations. We will first identify all the terms that can appear and at what order in $\epsilon$. We will solve the hydrodynamic equations with external sources to identify the Hall viscosity and angular momentum density in the frame where there is no current $J_3^i=0$,\footnote{Here we are referring to a physical frame and not to the ambiguity in the choice of hydrodynamic variables.} which we identify as the ground state of the system. We will match the hydrodynamic solutions with linear response to derive Kubo formulas that determine the transport coefficients responsible for the parity breaking physics. This will be useful later to compare with the holographic $p$-wave model.

In the linear response analysis we set to zero the velocity of the normal component, so the results are valid for both the locking in the lab frame and in the rest frame. When the velocity is non-zero but small there will be a Hall viscosity term in the stress tensor of the form
\begin{equation}
T_H^{ij} = -\widetilde{\eta}_H\epsilon^{(ik}\sigma^{j)}_k, \ \ \sigma_{ij}=\partial_i u_j+\partial_j u_i-\delta_{ij}\partial_k u^k.
\end{equation}
It turns out that the coefficient of this term depends on the type of locking. As we will see it is the same as the linear response coefficient for a locking in the lab frame but parametrically larger (close to the critical point)  for locking in the rest frame.

\subsection{Terms in the constitutive relations}\label{4a}
\label{sec:cons}

We list all the possible terms that appear in the locked phase in the Appendix \ref{sec:diss2} and \ref{sec:diss3}. For the locking in the lab frame, we list the terms that break parity and the order at which they appear in the expansion. We also give their approximate form for small velocities:
\begin{itemize}
\item Tensor $\tau^{\mu\nu}$:
\begin{equation}
\begin{split}
& O(\epsilon^2) \ \  \sigma^{\mu}_\nu\tzeta^{\alpha\nu}\sim \vev{\cO}^2\epsilon^{ik}\sigma_k^j.
\end{split}
\end{equation}
This term introduces the Hall viscosity.
\item Mixed tensor $\tau^\mu_A$:
\begin{equation}
\begin{split}
& O(\epsilon) \ \  \sigma^\mu_\alpha\tzeta^\alpha_A\sim \vev{\cO}\epsilon_{ a}^{\ k}\sigma_k^i.
\end{split}
\end{equation}
\item Vector $V^\mu$:
\begin{equation}
\begin{split}
& O(\epsilon^2) \ \   s_A\tzeta^\mu_A\sim  \vev{\cO}\epsilon^{ij}\partial_j\vev{\cO},\\
& O(\epsilon^2) \ \  v_{\alpha A}\tzeta^{\mu\alpha}\mu_A\sim \vev{\cO}^2\epsilon^{ij}\partial_j\left( \frac{\mu}{T}\right).
\end{split}
\end{equation}
The second term introduces a Hall conductivity $J_3^i =\sigma_H^{33}\epsilon^{ij} E_{3\,j}$.
\item Internal vector $V_{A}$:
\begin{equation}\label{scalar1}
\begin{split}
& O(\epsilon) \ \  s_B C_{BA}\sim \epsilon_a^{ \ i}\partial_i\vev{\cO},\\
& O(\epsilon) \ \   v_{\alpha B}\tzeta^\alpha_A\mu_B\sim \vev{\cO}\epsilon_a^{\ i}\partial_i\left( \frac{\mu}{T}\right).
\end{split}
\end{equation}
\item Scalar $\Sigma$: no terms.
\end{itemize}
Note, that because $s_a\sim \partial_a\vev{\cO}$, the associated terms are actually one order higher in $\epsilon$ than na{\"\i}vely expected.

For the locking in the rest frame we have the same terms, plus a few additional more. All the new terms are proportional to $v^\mu_A$ as given in \eqref{bbslock2}, and they come from the term $\propto P^{\mu\alpha}\partial_\alpha u_A$. We can decompose the transverse derivative of the velocity in shear, curl and scalar components:
\begin{equation}
P_\mu^{\alpha}\partial_\alpha u_A=\half \sigma_{\mu A}+\half \omega_{\mu A}+\half P_{\mu A}\theta.
\end{equation}
At small velocities we do the approximation
\begin{equation}
P_\mu^{\alpha}\partial_\alpha u_A\simeq \half \sigma_{ij}\delta^j_a+\half \omega_{i j}\delta^j_a+\half\delta_{i a}\partial_k v^k,
\end{equation}
where $\omega_{ij}=\partial_i u_j-\partial_j u_i$ is the vorticity.
The extra terms are then
\begin{itemize}
\item Tensor $\tau^{\mu\nu}$:
\begin{equation}
\begin{split}
& O(\epsilon) \ \ v^\mu_A\tzeta^\nu_A\sim \vev{\cO}\epsilon^{ik}(\sigma_k^j+\omega_k^{\ j}+\delta_k^j\theta).
\end{split}
\end{equation}
The last term $\propto \theta$ actually drops from the symmetric stress tensor, while the second term $\propto \omega_{ij}$ is a scalar contribution, as one can check by using $\omega_{ij}=\epsilon_{ij}\omega$.
\item Mixed tensor $\tau^\mu_A$:
\begin{equation}
\begin{split}
& O(1) \ \  v^\mu_B C_{BA}\sim \epsilon_a^{\ k}(\sigma_k^j+\omega_k^{\ j}+\delta_k^j\theta),\\
& O(\epsilon^2) \ \  v_{\alpha\, B}\zeta^{\mu\alpha}C_{BA},\ v_{\alpha B}\tzeta^{\mu\alpha}P_{BA},   v_{\alpha\,B}\hzeta^\mu_B\tzeta^\alpha_A,\  v_{\alpha\,B}\tzeta^\mu_B\hzeta^\alpha_A\\
&\sim \vev{\cO}^2\epsilon_a^{\ k} (\sigma_k^j+\omega_k^{\ j}+\delta_k^j\theta).
\end{split}
\end{equation}
\item Vector $V^\mu$: no additional terms.
\item Internal vector $V_{A}$: no additional terms.
\item Scalar $\Sigma$:
\begin{equation}\label{scalar2}
\begin{split}
& O(\epsilon) \ \   v_{\alpha A}\tzeta^\alpha_A\sim \vev{\cO}\epsilon^{ij}\omega_{ij}=2\vev{\cO}\omega.
\end{split}
\end{equation}
\end{itemize}

There are two main observations we wish to make. The first is that locking does not necessarily introduce all possible parity breaking terms. For instance, terms depending on the vorticity are absent for the locking in the lab frame, and terms depending on the magnetic field are absent in both cases. For comparison, a complete list of parity breaking terms in normal fluids can be found in \cite{Jensen:2011xb}.\footnote{A similar study for non-relativistic fluids was made in \cite{Kaminski2013}.}
The second is that terms depending on the strain rate $\sigma^{ij}$ (tensor) are parametrically larger when we do the locking in the rest frame. On the other hand, the terms depending on gradients of chemical potential and the expectation value $\vev{\cO}$ (vector) are the same for both lockings. The last are responsible for the angular momentum density, so we find that
\begin{equation}
\frac{\widetilde{\eta}_H^{\rm lab}}{\ell}\sim 1, \ \ \frac{\widetilde{\eta}_H^{\rm rest}}{\ell}\sim \frac{1}{\epsilon}\gg 1.
\end{equation}
However, the Hall viscosity as computed from linear response $\eta_H$ is not the same as $\widetilde{\eta}_H$ for the locked phase in the rest frame. We will derive Kubo formulas for both Hall viscosities and for the angular momentum density in the following.

\subsection{Response to external metric and viscosities}\label{4b}

We introduce a background metric of the form $g_{00}=-1$, $g_{i0}=0$, $g_{ij}=\delta_{ij}+h_{ij}(t)$. The only contributions to dissipative terms that are of linear order in the metric come from the strain rate tensor
\begin{equation}
\sigma_{\mu\nu} = -2\Gamma^\alpha_{\mu\nu}u_\alpha.
\end{equation}
The non-zero components are space-like $\sigma_{ij}=\partial_t h_{ij}$. Using the results from the Appendix \ref{sec:diss1} we get the following dissipative terms to $O(\epsilon^2)$
\begin{itemize}
\item Tensor $\tau^{\mu\nu}$
\begin{equation}
\begin{split}
& O(1) \ \ \sigma^{ij},\\
& O(\epsilon^2) \ \  \vev{\cO}^2\sigma^{ij}, \  \vev{\cO}^2\sigma^{i}_k\epsilon^{kj}.
\end{split}
\end{equation}
\item Mixed tensor $\tau^\mu_A$
\begin{equation}
\begin{split}
& O(\epsilon) \ \ \vev{\cO}\sigma^i_a,\ \vev{\cO}\epsilon_{ak}\sigma^{ik}\\
\end{split}
\end{equation}
\item Vector $V_{1,2}^\mu$: none  up to $O(\epsilon^2)$.
\item Internal vector $V_{2\,A}$: none up to $O(\epsilon^2)$.
\item Scalar $\Sigma_{2,4,T}$: none up to $O(\epsilon^2)$.
\end{itemize}
The stress tensor and the currents become
\begin{equation}
\begin{split}
& T^{ij}=p_t \delta^{ij}-p_t h^{ij}-\eta \sigma^{ij}-2\eta_H \epsilon^{(ik}\sigma_k^{j)},\\
& J^i_a=\vev{\cal O}\delta^i_a-\kappa\sigma^i_a-\kappa_H \epsilon_{ak}\sigma^{ik},
\end{split}
\end{equation}
where
\begin{equation}
p_t=p+\frac{\vev{\cO}^2}{f_0}.
\end{equation}
This implies that the correlation functions with the traceless component of the stress tensor $T_{<ij>}\equiv T_{ij}-\frac{1}{2}\left((\delta_{ij}+h_{ij}) T^k_k-\delta_{ij}h^{kl}T_{kl}\right)$ are
\begin{align}
&\vev{T^{<ij>} T^{<kl>}}=i\omega\frac{\eta}{2} \left(\delta^{ik}\delta^{jl} +\delta^{il}\delta^{jk}-\delta^{ij}\delta^{kl}\right) +i\omega\frac{\eta_H}{2}\left(\epsilon^{ik}\delta^{jl} +\epsilon^{jk}\delta^{il}+\epsilon^{il}\delta^{jk}+\epsilon^{jl}\delta^{ik}\right),\\
&\vev{J^i_a T^{<kl>}}= i\omega \frac{\kappa}{2}\left(\delta^{ik}\delta^{l}_a +\delta^{il}\delta^{k}_a-\delta^{i}_a\delta^{kl}\right) +i\omega\frac{\kappa_H}{2}\left(\epsilon_a^{\ k}\delta^{il}+\epsilon_a^{\ l}\delta^{ik}-\epsilon_a^{\ i}\delta^{kl}\right).
\end{align}
The Kubo formulas for the parity-breaking coefficients are
\begin{align}
&\eta_H=\lim_{\omega\to 0}\frac{1}{4i\omega}\epsilon_{ik}\delta_{jl}\vev{T^{<ij>} T^{<kl>}}(\omega,\mathbf{k}=0),\\
&\kappa_H=\lim_{\omega\to 0}\frac{1}{2i\omega}\epsilon_{ik}\delta_{l}^a\vev{J^i_a T^{<kl>}}(\omega,\mathbf{k}=0).
\end{align}
Similar Kubo formulas for the Hall viscosity were derived in \cite{Saremi:2011ab,Jensen:2011xb,Bradlyn2012}.

\subsection{Angular momentum density}\label{4c}

The equilibrium solution in the locked phase has a finite normal density $q_3=q_n$ and the following values for the superfluid velocities:
\begin{equation}
\xi_{3\,0}=\mu, \ \ \xi_{a\,i}=-\frac{\vev{\cO}}{f_0}.
\end{equation}
We now allow the temperature, chemical potential and $\vev{\cO}$ to vary slowly over space, but keeping a static configuration and zero velocity for the normal component. In the absence of sources
\begin{equation}
T^{0i}=-q_s \delta\xi^i_3, \ \ J_3^i=-\frac{q_s}{\mu}\delta\xi^i_3+\nu^i_3,
\end{equation}
where we are expanding only up to first order terms. We are interested in configurations where the current vanishes $J_3^i=0$, so that $\delta\xi^i_3=\frac{\mu}{q_s}\nu^i_3$ and
\begin{equation}
T^{0i}=-\mu\nu^i_3.
\end{equation}
The non-zero dissipative terms are proportional to
\begin{equation}
v^i_3= \partial^i\left(\frac{\mu}{T}\right) , \ \ s_a=\partial_a\vev{\cal O}.
\end{equation}
Then, we have the following independent contributions to $T^{0i}$:
\begin{equation}
\nu_3^i=-C_1 v^i_3-C_2 s^i-C_3 \epsilon^{ij}v_{3\,j}-C_4 \epsilon^{ij} s_j.
\end{equation}
All the contributions are $O(\epsilon^2)$. Let us assume the coefficients $C_n$ are approximately constant, then the total angular momentum is, for a smoothly changing condensate
\begin{equation}
L_{\rm smooth}=\int d^2 x\,\epsilon_{ij}x^i T^{0j}=\int d^2 x  ~2\mu \left( C_3 \frac{\mu}{T} + C_4 \vev{\cal O}\right).
\end{equation}
If we have a `droplet' of superfluid of radius $r_0$ with constant density and condensate (so for instance $\vev{\cal O}\sim \vev{\cal O}\Theta(r_0-r)$ and similarly for $\mu$), the angular momentum picks up a contribution from the boundary of the droplet:
\begin{equation}
L_{\rm droplet}=2\mu \pi r_0^2\left( C_3 \frac{\mu}{T} + C_4\vev{\cal O}\right).
\end{equation}
Therefore, we can define the average angular momentum density as
\begin{equation}
\ell=\frac{L_{\rm droplet}}{\pi r_0^2}=  2\mu\left( C_3 \frac{\mu}{T} + C_4 \vev{\cal O} \right).
\end{equation}
Since the contributions are of order $\ell \sim \vev{\cal O}^2\sim \epsilon^2$, we found that generically
\begin{equation}
\frac{\eta_H}{\ell}\sim  1.
\end{equation}

\subsubsection{Kubo formulas}

In order to find the Kubo formulas for the angular momentum density we will need to solve the hydrodynamic equations to leading order in derivatives and linear order in the sources. We set the velocity to constant $u^\mu=(1,0)$ and consider only static configurations. As external source we will allow only a constant gauge potential $A_{a\,0}$, and we will allow a fluctuation $\delta\xi_{0\,3}$.

From the current conservation equation we get
\begin{equation}
0=D_\mu J^\mu_A=u^\mu D_\mu q_A-D_\mu(f_{AB} \xi^\mu_B) \ \ \Rightarrow s_A=u^\mu D_\mu q_A=D_0 q_A.
\end{equation}
Then,
\begin{equation}
s_3=0, \ \ s_a=q_n \epsilon_{ab}A_{b\,0}.
\end{equation}
The Josephson condition $\xi_{3\,0}=\mu_3$ together with the equation for the superfluid velocity $D_\mu \xi_{A\,\nu}-D_\nu\xi_{A\,\mu}=0$ leads to
\begin{equation}
D_i\mu_3=D_0\xi_{3\,i}=\frac{\vev{\cO}}{f_0}\epsilon_{ia}A_{a\,0}.
\end{equation}
We will now use the conservation equation of the stress tensor $\partial_k T^{ki}=0$
\begin{equation}
B_T\partial_i  T+B_\mu D_i\delta \mu_3=0,
\end{equation}
where
\begin{equation}
B_X=\frac{\partial p}{\partial X}+\frac{\vev{\cO}^2}{f_0^2}\frac{\partial f_0}{\partial X}, \ \ X=T,\mu.
\end{equation}
The derivatives are evaluated at constant $\zeta^2=2\frac{\vev{\cO}^2}{f_0^2}$. Combining everything, we find $v^i_a=0$ and
\begin{equation}\label{Qdef}
v^i_3=\left(1+\frac{\mu B_\mu}{T B_T} \right)\frac{\vev{\cO}}{f_0 T}\epsilon_{ia}A_{a\,0}\equiv Q\epsilon_{ia}A_{a\,0}.
\end{equation}

Then, the current becomes
\begin{equation}
J^i_3=-(C_1Q+C_2q_n)\epsilon^i_{\ b}A_{b\,0}+(C_3Q+C_4 q_n) \delta^{ib}A_{b\,0}.
\end{equation}
We obtain the following Kubo formulas, for the correlators evaluated at $\omega=0$, $\mathbf{k}=0$,
\begin{equation}
\begin{split}
&C_1Q+C_2q_n=-\frac{1}{2}\epsilon_{ia}\vev{J^i_3 J^0_a},\\
&C_3Q+C_4q_n=\frac{1}{2}\delta_{ia}\vev{J^i_3 J^0_a}.
\end{split}
\end{equation}
We now introduce a non-zero space-dependent potential $A_{3\,0}$, so the electric field is non-zero $E_{3\,i}=\partial_i A_{3\,0}\neq 0$ and allow $\delta\xi_{a\,0}$ to fluctuate. The equations for current conservation imply that $s_A=0$.
The Josephson condition $\delta\xi_{a\,0}=\delta\mu_a$ together with the equation for the superfluid velocity $D_\mu \xi_{A\,\nu}-D_\nu\xi_{A\,\mu}=0$ leads to
\begin{equation}
D_i\mu_a=D_0\xi_{a\,i}=\frac{\vev{\cO}}{f_0}\epsilon_{ai}A_{3\,0}.
\end{equation}
We will now use the conservation equation of the stress tensor $\partial_k T^{ki}=qE^i_3$
\begin{equation}\label{solT}
B_T\partial_i  T+B_{\mu_a} D_i\delta \mu_a=qE_{3\,i}.
\end{equation}
The derivatives are evaluated at constant $\zeta^2=2\frac{\vev{\cO}^2}{f_0^2}$ and $\mu_3$. Combining everything, we find
\begin{equation}\label{sola30}
\begin{split}
v_{a\,i} &= \frac{\vev{\cO}}{f_0 T}\epsilon_{ai}A_{3\,0},\\
v_{3\,i} &=-\left(1+\frac{q\mu}{TB_T} \right)\frac{E_{3\,i}}{T}+\frac{\mu B_{\mu_a}}{TB_T}\frac{\vev{\cO}}{f_0T}\epsilon_{ai}A_{3\,0}\equiv -\alpha_0 \partial_i A_{3\,0}+\alpha_i A_{3\,0}.
\end{split}
\end{equation}

We are left with the current
\begin{equation}
J^i_3 = C_1\alpha_0 \partial^i A_{3\,0}-C_1\alpha^i A_{3\,0}+C_3\alpha_0\epsilon^{ij}\partial_j A_{3\,0}-C_3\epsilon^{ij}\alpha_j A_{3\,0}.
\end{equation}
Therefore,
\begin{equation}\label{c3}
C_3= -\frac{i}{2\alpha_0}\lim_{\mathbf{k}\to 0}\epsilon_{ij}\frac{\partial }{\partial k_j}\vev{J^i_3 J^0_3}(\omega=0,\mathbf{k}).
\end{equation}
Note that $C_3$ is related to the Hall conductivity, we will comment more on this in the conclusions.

\subsection{Hall viscosity term at non-zero velocity}\label{4d}

The enhanced Hall viscosity  that appears when we do the locking in the rest frame is generated by a term depending on $v_a^i$ in $\pi^{ij}$. Note that the equation of motion for the superfluid velocity and the Josephson condition imply
\begin{equation}
D_i\mu_a=E_{a\,i}+D_0\xi_{a\,i}.
\end{equation}
Therefore,
\begin{equation}
v_{a\,i}=\frac{1}{T}D_0\xi_{a\,i}.
\end{equation}
We introduce a source $A_{a\,i}$ which is time-dependent but independent of the spatial coordinates and satisfies the conditions
\begin{equation}\label{condongauge}
\delta^i_aA_{a\,i}=0, \ \ \epsilon^i_{\ a}A_{a\,i}=0.
\end{equation}
The hydrodynamic equations are automatically satisfied to leading order in derivatives and the sources. The Fourier transform of the superfluid velocity is
\begin{equation}
\delta\xi_{a\,i}= A_{a\,i}+\frac{\mu}{i\omega}\epsilon_{ab}A_{a\,i}.
\end{equation}
The stress tensor including the first order dissipative terms has the form
\begin{equation}
T^{ij}=p_t\delta^{ij}-\vev{\cO}(\delta^i_a\delta\xi_a^j+\delta^j_a\delta\xi^i_a)-C_\eta \delta^{(i}_a v_a^{j)}-C_{\eta_H}\epsilon^{(i}_{\ a}v_a^{j)}.
\end{equation}
Then, taking the variation with respect to the gauge field
\begin{equation}
\begin{split}
\vev{T^{<ij>} J^k_a} &=-\left[\vev{\cO}+\frac{i\omega C_\eta}{2T}-\frac{\mu C_{\eta_H}}{2T} \right](\delta^i_a\delta^{jk}+\delta^j_a\delta^{ik}-\delta^{ij}\delta_k^a)\\
&- \left[\frac{\mu\vev{\cO}}{i\omega}+ \frac{\mu C_\eta}{2T}+\frac{i\omega C_{\eta_H}}{2T} \right](\epsilon^i_{\ a}\delta^{jk}+\epsilon^j_{\ a}\delta^{ik}-\delta^{ij}\epsilon^k_{\ a}).
\end{split}
\end{equation}
Note, that the conditions \eqref{condongauge} for the gauge field are satisfied by the correlator
\begin{equation}
\vev{T^{<ij>} J^k_a}\delta^k_a=\vev{T^{<ij>} J^k_a}\epsilon_{ka}=0.
\end{equation}
The Hall viscosity coefficient is related to $C_{\eta_H}$ as
\begin{equation}
\widetilde{\eta}_H=\frac{\mu}{2T}C_{\eta_H}.
\end{equation}
It is straightforward to derive the following Kubo formula
\begin{equation}
\widetilde{\eta}_H=\vev{\cO}+\frac{1}{4}\lim_{\omega\to 0}\delta_{ia}\delta_{jk}\vev{T^{<ij>} J^k_a}(\omega,\mathbf{k}=0).
\end{equation}

\section{Comparison with holographic $p$-wave model}
\label{sec:holo}

In this section we will check the consistency of the general hydrodynamic analysis by comparing with the results obtained by Son and Wu \cite{Son:2013xra} for the angular momentum density and Hall viscosity in the holographic $p$-wave model \cite{Gubser:2008zu,Gubser:2008wv,Roberts:2008ns}. We also compute the rest frame Hall viscosity $\widetilde{\eta}_H$ and find that the leading order contribution actually vanishes in this model. In the following
we present the basic features of the model and the results. We have collected the equations of motion and useful formulas in Appendix~\ref{sec:eqs}.

The holographic $p$-wave model consists of Einstein gravity plus a cosmological constant coupled to a non-Abelian $SU(2)$ gauge field in 3+1 dimensions.
\begin{equation}
S=\frac{1}{2\kappa^2}\int d^4 x\,\sqrt{-g}\,\left(R-2\Lambda-\frac{1}{4}F^A_{\mu\nu}F^{A\,\mu\nu} \right).
\end{equation}
The background metric and gauge field are charged black hole solutions of the form
\begin{equation}
\begin{split}
&ds^2=-F(z) dt^2+\frac{dz^2}{F(z)}+r^2(z)(dx^2+dy^2),\\
&A^3_{0}=\phi(z), \ \ A^a_i=A(z)\delta^a_{i}.
\end{split}
\end{equation}
As $z\to \infty$ the metric approaches asymptotically $AdS_4$. Applying the holographic dictionary, this means that the dual field theory is a CFT with a $SU(2)$ global symmetry at a finite density and finite temperature state. There can also be zero temperature black holes but we will not discuss them here.

The solutions for the gauge field are such that, as $z\to \infty$,\footnote{This amounts to taking $R=1/2$ in \cite{Son:2013xra}.}
\begin{equation}
\phi\simeq \mu+\frac{\vev{J^3_0}}{z}, \ \ A\simeq \frac{\vev{\cO}}{z}.
\end{equation}
From the dual field theory point of view this means that there is a non-zero chemical potential $\mu_3=\mu$ and an expectation value for the current $\vev{J_a^i}=\vev{\cO}\delta^i_a$. Therefore, the $SU(2)$ charged black holes presented above describe a locked phase like the ones we have analyzed using hydrodynamics. This model has a second order phase transition at a critical temperature $T_c$ from the locked phase to an unbroken phase. Close to the critical point
\begin{equation}
\vev{\cO}\sim \sqrt{(T_c-T)/T_c}\sim \epsilon,
\end{equation}
and we can apply the same expansion that we used in the hydrodynamic model. The near-critical expansion was used in \cite{Son:2013xra} to compute the values of the Hall viscosity and angular momentum density, so we can make a direct comparison.

\subsection{Correlators and Kubo formulas}

Let us collect here the expected orders in $\epsilon$ from the hydrodynamic analysis:
\begin{itemize}
\item Hall viscosity $\eta_H\sim O(\epsilon^2)$:
\begin{equation}
\eta_H=\lim_{\omega\to 0}\frac{1}{4i\omega}\epsilon_{ik}\delta_{jl}\vev{T^{<ij>} T^{<kl>}}(\omega,\mathbf{k}=0).
\end{equation}
\item Angular momentum density $\ell\sim O(\epsilon^2)$:
\begin{equation}
\begin{split}
&\ell=2\mu C_3 \frac{\mu}{T} +2\mu C_4 \vev{\cal O},\\
&C_3Q+C_4q_n=\frac{1}{2}\delta_{ia}\vev{J^i_3 J^0_a}(\omega=0,\mathbf{k}=0),\\
&C_3= -\frac{i}{2\alpha_0}\lim_{\mathbf{k}\to 0}\epsilon_{ij}\frac{\partial }{\partial k_j}\vev{J^i_3 J^0_3}(\omega=0,\mathbf{k}).
\end{split}
\end{equation}
Where $Q$ is defined in \eqref{Qdef} and $\alpha_0$ in \eqref{sola30}.
\item Hall viscosity for locking in the rest frame $\widetilde{\eta}_H\sim O(\epsilon)$:
\begin{equation}\label{wetah}
\widetilde{\eta}_H=\vev{\cO}+\frac{1}{4}\lim_{\omega\to 0}\delta_{ia}\delta_{jk}\vev{T^{<ij>} J^k_a}(\omega,\mathbf{k}=0).
\end{equation}
\end{itemize}
There are four correlators whose leading order in $\epsilon$ we need to estimate. The calculation of $\eta_H$ using the Kubo formula was made in \cite{Son:2013xra} originally. One can use their result to show that it is $O(\epsilon^2)$, but in Appendix \ref{sec:corr} we present a derivation that makes it more explicit. The angular momentum density was also computed in \cite{Son:2013xra}, but using a different method. It would be interesting to compare the exact (numerical) value obtained from the Kubo formula with their result, but here we will limit ourselves to an estimation of the order of magnitude.

The correlators of the energy-momentum tensor and global $SU(2)$ current in the dual field theory can be computed by evaluating the (properly renormalized) on-shell action of small fluctuations around the background solution. The fluctuations of the metric and gauge field take the form
\begin{equation}
g_{\mu\nu}= \bar{g}_{\mu\nu}+h_{\mu\nu}, \ \ A_\mu^A =\bar{A}_\mu^A+a_\mu^A ,
\end{equation}
where $\bar{g}_{\mu\nu}$ and $\bar{A}^A_{\mu}$ are the background solutions. We perform an expansion of the equations of motion in $A\sim \epsilon$ and solve the equations order by order. In most cases we will not need to find the explicit form of the solution to estimate the order of the transport coefficients.

Following \cite{Son:2013xra}, at zero momentum we can split the fluctuations according to their representation under the unbroken $U(1)$ group mixing space and time components. For the metric and gauge field they group into tensor, vector and scalar. Both Hall viscosity coefficients appear from tensor fluctuations, while the angular momentum density has a contribution from the vector fluctuation and a contribution that originates from momentum-dependent fluctuations that mix scalar with vector fluctuations.

These are the fluctuations that we will turn on in order to compute each of the coefficients:
\begin{itemize}
\item Hall viscosity $\eta_H$: time-dependent $h_{ij}$ with $\delta^{ij} h_{ij}=0$.
\item Angular momentum density: a constant vector contribution $a_i^3$ and $a_0^a$, and a space-dependent mixed contribution $a_i^3$, $a_0^3$.
\item Hall viscosity in rest frame $\widetilde{\eta}_H$: time-dependent $h_{ij}$ with $\delta^{ij} h_{ij}=0$ and $a_i^a$ with $\delta^i_a a_i^a=0$.
\end{itemize}
The expansion of the fluctuations close to the boundary $z\to\infty$ is
\begin{equation}
h_{ij}\simeq {\cH}_{ij}+\frac{\cT_{ij}}{z^3}+\cdots, \ \ \ a_\mu^A=\cA_\mu^A+\frac{\cJ^A_\mu}{z}+\cdots.
\end{equation}
Where in the dual field theory ${\cH}_{ij}$ and $\cA_\mu^A$ are the sources for the energy-momentum tensor and global $SU(2)$ current and $\cT_{ij}$ and $\cJ^A_\mu$ are proportional to the expectation values, following the usual $AdS/CFT$ dictionary. The correlators are found by taking variations of the on-shell action with respect to the sources
\begin{equation}
\begin{split}
&\vev{J_A^\mu J_B^\nu}=\frac{\delta^2 S_{\text{on-shell}}}{\delta \cA_{\mu}^A\delta \cA_\nu^B},\\
&\vev{T^{\mu\nu}J_A^\lambda}=\frac{\delta^2 S_{\text{on-shell}}}{\delta \cH_{\mu\nu}\delta \cA_\lambda^A},\\
&\vev{T^{\mu\nu}T^{\lambda\sigma}}=\frac{\delta^2 S_{\text{on-shell}}}{\delta \cH_{\mu\nu}\delta \cH_{\lambda\sigma}}.
\end{split}
\end{equation}
The details about the renormalization of the on-shell action and the derivation of the equations of motion can be found in the original reference \cite{Son:2013xra}. In the following we estimate the order of the transport coefficients.

\subsection{Vector contribution to angular momentum density}

For a vector fluctuation with $h_{0i}=0$, $a_i^3$ and $a_0^a$, the equations of motion are, to leading and next-to-leading order in $A$\footnote{The full equations can be found in the Appendix \eqref{veceqs} and \eqref{veceqs2}.}
\begin{equation}
\begin{split}
& r^2 \phi' (a_i^3)'=-r^2 A'(a_0^i)',\\
& [F(a_i^3)']'=\frac{\lambda^2\phi A}{F}a_0^i,\\
& \left(\frac{a_0^i}{\phi} \right)' = \frac{F}{r^2\phi^2}\left[A' a_i^3-A(a_i^3)' \right].
\end{split}
\end{equation}
To leading order in $A$ we have the solutions
\begin{equation}
a_i^3= {\cal A}_i^3, \ \ a_0^a = \frac{\phi}{\mu}{\cal A}_0^a.
\end{equation}
To next order in $A$ the solutions become
\begin{equation}
\begin{split}
a_i^3 &= {\cal A}_i^3-\frac{A}{\mu}{\cal A}_0^i,\\
a_0^a &= \frac{\phi}{\mu}{\cal A}_0^a-{\cal A}_i^3\phi\int_z^\infty d\tilde z\, \frac{F(\tilde z) A'(\tilde z)}{r^2(\tilde z)\phi^2(\tilde z)}.
\end{split}
\end{equation}
The on-shell action is
\begin{equation}
S=-\lim_{z\to\infty} \frac{1}{4\kappa^2}\int d^3x\, \sqrt{-\bar{g}}a_\nu^A f^{A\, z\nu}.
\end{equation}
Since we are in the gauge where the radial components are zero, $f^A_{z\mu}=\partial_z a_\mu^A$. Then, the action becomes
\begin{equation}
S=-\lim_{z\to\infty} \frac{1}{4\kappa^2}\int d^3x\, r^2\left[-a_0^a\partial_z a_0^a+\frac{F}{r^2}a_i^3\partial_z a_i^3 \right].
\end{equation}
In particular, the cross term is
\begin{equation}
S_{\rm cross}=\lim_{z\to\infty} \frac{1}{2\kappa^2}\int d^3x\, {\cal A}_0^a{\cal A}_a^3\frac{1}{\mu}\left[FA'-r^2\phi'\phi\int_z \frac{F A'}{r^2\phi^2} \right].
\end{equation}
The expansion close to the boundary is
\begin{equation}
F\simeq 4z^2, \ \ \phi \simeq \mu, \ \ \phi' \simeq -\frac{\phi_1}{z^2}, \ \ A\simeq \frac{\alpha_1}{z}, \ \ r^2\simeq 4 z^2.
\end{equation}
This gives
\begin{equation}
S_{\rm cross}=\frac{1}{2\kappa^2}\int d^3x\, {\cal A}_0^a{\cal A}_a^3\left[\frac{-4\alpha_1}{\mu}\right].
\end{equation}
Using that the expectation value of the dual current is
\begin{equation}
\vev{{\cal O}}= \frac{2\alpha_1}{\kappa^2},
\end{equation}
we find that the two-point function to leading order in the vev and at zero frequency and momentum is
\begin{equation}
\vev{J^i_3 J^{0}_a}(\omega=0,\mathbf{k}=0)=-\frac{\vev{{\cal O}}}{\mu}\delta_a^{i}.
\end{equation}
Then,
\begin{equation}\label{c4val}
\delta_i^a\vev{J^i_3 J^{0}_a}(\omega=0,\mathbf{k}=0)\sim \epsilon.
\end{equation}

\subsection{Mixed contribution to the angular momentum density}

We want to check what is the order of the $\vev{J_0^3 J_i^3}\sim \epsilon_{ij}k_j$ contribution. We will do it in the probe approximation, where the metric fluctuation is set to zero by hand. The equations of motion of the gauge fluctuation $a_\mu^A$ in the background $A_\mu^A$ are
\begin{equation}
\nabla_\mu f^{A\,\mu\nu}+\lambda \epsilon^{ABC}(A_\mu^B f^{C\,\mu\nu}+a_\mu^B F^{C\,\mu\nu})=0.
\end{equation}
Assuming time derivatives are zero, this becomes
\begin{equation}
\begin{split}
&0 = \partial_z f^{A\,z\nu}+\partial_if^{A\,i\nu}+2\frac{r'}{r}f^{A\,z\nu} \\
&+\lambda\left[ \epsilon^{A3C}\phi f^{C\,0\nu}+\epsilon^{AiC}A f^{C\,i\nu}+\delta^\nu_z\phi'\epsilon^{Ab3}a_0^b-\delta^\nu_z\epsilon^{ABi}\frac{A'F}{r^2}a_i^B
- \epsilon^{ABC} \epsilon^{C3i} \frac{\phi A}{Fr^2} (a_0^B \delta_i^\nu - a_i^B \delta_0^\nu) \right].
\end{split}
\end{equation}
We are working in the gauge $a_z^A=0$ and impose the condition $\partial_i a_i^3=0$. We can set $a_i^a=0$, the $A=3,\nu=z$ equation is automatically satisfied. The $A=a$, $\nu=z$ equation is
\begin{equation}
\epsilon^{ab}(\phi (a_0^b)'-\phi' a_0^b)-\epsilon^{ai}\frac{F}{r^2}(A(a_i^3)'-A' a_i^3)=0.
\end{equation}
The solution to this equation is
\begin{equation}
a_0^b=\delta^{ib}\phi\int^z \frac{F}{r^2}(A(a_i^3)'-A' a_i^3).
\end{equation}
The remaining $\nu\neq z$ equations are
\begin{equation}
\begin{split}
&\partial_z f^{A\,z0}+\partial_i f^{A\,i0}+2\frac{r'}{r}f^{A\,z0}+\lambda \left(A \epsilon^{AiC}f^{C\,i0}+ \epsilon^{ABC} \epsilon^{C3i} \frac{\phi A}{Fr^2} a_i^B\right) =0,\\
&\partial_z f^{A\,zj}+\partial_i f^{A\,ij}+2\frac{r'}{r}f^{A\,zj}+\lambda \left(\epsilon^{A3C}\phi f^{C\,0j}+\epsilon^{AiC}Af^{C\,ij}- \epsilon^{ABC} \epsilon^{C3i} \frac{\phi A}{Fr^2} a_0^B \right)=0.
\end{split}
\end{equation}
Where the field strengths are
\begin{equation}
\begin{split}
&f^A_{z\mu}=\partial_z a_\mu^A,\ \ f^3_{i0}=\partial_i a_0^3, \ \ f^3_{ij}=\partial_i a_j^3-\partial_j a_i^3,\\
&f^a_{i0}=\lambda A\epsilon^{ai}a_0^3, \ \ f^a_{ij}=\lambda A(\epsilon^{ai}  a_j^3-\epsilon^{aj}a_i^3).
\end{split}
\end{equation}
We can rewrite the equations as
\begin{equation}
\begin{split}
f^{A\,z0} &=-\frac{1}{r^2}\int^zdz\left(\partial_i f^{A\,i0}+\lambda \left(A \epsilon^{AiC}f^{C\,i0} +  \epsilon^{ABC} \epsilon^{C3i} \frac{\phi A}{Fr^2} a_i^B\right)\right),\\
f^{A\,zi} &=-\frac{1}{r^2}\int^zdz\left(\partial_i f^{A\,ij}+\lambda \left(\epsilon^{A3C}\phi f^{C\,0j}+\epsilon^{AiC}Af^{C\,ij}- \epsilon^{ABC} \epsilon^{C3i} \frac{\phi A}{Fr^2} a_0^B\right)\right).
\end{split}
\end{equation}
Then, the on-shell action will contain terms of the form
\begin{equation}
\begin{split}
S &=-\frac{1}{4\kappa^2} \int_{z\to\infty} d^3x\, \sqrt{-g}a_\mu^A f^{A\,z\mu}\\
&=\frac{1}{4\kappa^2}\int_{z\to\infty} d^3x\,\left[a_0^3\int^zdz\left(\partial_i f^{3\,i0}+\lambda A \epsilon^{ia}f^{a\,i0} \right)\right.\\&\left.
+a_i^3\int^zdz\left(\partial_i f^{3\,ij}+\lambda \left(\epsilon^{ia}Af^{a\,ij}- \frac{\phi A}{Fr^2} a_0^i\right)\right)\right.\\&\left.+a_0^a\int^zdz\left(\partial_i f^{a\,i0}+\lambda \left( A \epsilon^{ai}f^{3\,i0}- \frac{\phi A}{Fr^2} a_i^3 \right)\right) \right].
\end{split}
\end{equation}
The first two terms don't mix the $a_i^3$ and $a_0^3$ fluctuations. The first two in the last term are roughly
\begin{equation}
\sim A^2 a_i^3 \epsilon^{ij}\partial_j a_0^3.
\end{equation}
Therefore,
\begin{equation}
\vev{J^i_3 J^0_3}(\omega=0,\mathbf{k})\sim  i\epsilon^{ij}k_j A^2,
\end{equation}
and
\begin{equation}
\lim_{\mathbf{k}\to 0}\epsilon_{ij}\frac{\partial }{\partial k_j}\vev{J^i_3 J^0_3}(\omega=0,\mathbf{k}) \sim \epsilon^2.
\end{equation}
Together with \eqref{c4val}, this confirms the hydrodynamic analysis:
\begin{equation}
C_3\sim \epsilon^2,  \ \ C_4\sim \epsilon \ \  \Rightarrow \ \ \ell\sim \epsilon^2.
\end{equation}

\subsection{Hall viscosity in the rest frame}

To leading order in $A$, the equations for the tensor modes are\footnote{The full equations can be found in the Appendix \eqref{tenseq} and \eqref{tenseq2}.}
\begin{equation}
\begin{split}
&0= (r^2  F h_i')',\\
&0= (F a_i')'+\frac{\lambda^2\phi^1}{F}a_i.
\end{split}
\end{equation}
We are following the notation of \cite{Son:2013xra}, where $h_i=\{h_{xy}, h_{xx}-h_{yy}\}$, $a_i=\{a_x^2+a_y^1 ,a_x^1-a_y^2 \}$.

The solution regular at the horizon for the metric is simply the constant solution $h_i=h_i^b$. For the gauge field it takes the form
\begin{equation}
a_i=\frac{A^{(1)}}{\alpha_0^{(1)}} {\cal A}_i^b,
\end{equation}
where $A^{(1)}$ is the regular solution of the background equations of motion when they are linearized in $A$. It asymptotes the value  $\alpha_0^{(1)}$ as $z\to \infty$.

To next-to-leading order, we have to solve the equations
\begin{equation}
\begin{split}
&0= (r^2  F \delta h_i')'-2\left[-F A' {A^{(1)}}'+\frac{\lambda^2\phi^2 A}{F}A^{(1)} \right]\frac{{\cal A}_i^b}{\alpha_0^{(1)}},\\
&0= (F \delta a_i')'+\frac{\lambda^2\phi^2}{F}\delta a_i.
\end{split}
\end{equation}
We can make $\delta a_i=0$. Using the equation of motion for $A^{(1)}$, we can simplify the equation for $\delta h_i$ to
\begin{equation}
0= (r^2  F \delta h_i')'+2(F A{ A^{(1)}}')'\frac{{\cal A}_i^b}{\alpha_0^{(1)}},
\end{equation}
Then, the solution is
\begin{equation}
\delta h_i = 2\frac{{\cal A}_i^b}{\alpha_0^{(1)}}\int_z^\infty d\tilde z \frac{A {A^{(1)}}'}{r^2}.
\end{equation}
Note that the solution is regular at the horizon. The expansion close to the boundary is
\begin{equation}
\begin{split}
& h_i \simeq h_i^b +2\frac{{\cal A}_i^b}{\alpha_0^{(1)}}\times O\left(\frac{1}{z^4}\right),\\
& a_i \simeq {\cal A}_i^b\left( 1+ \frac{\alpha_1^{(1)}}{\alpha_0^{(1)}}\frac{1}{z}+\cdots\right).
\end{split}
\end{equation}
To this order, there is only a mixed contribution to the on-shell action coming from the term
\begin{equation}
S^{(2)}_{\rm mix}=\lim_{z\to\infty} \frac{1}{2\kappa^2}\int d^3x\,FA' h_i a_i=\int d^3x\,\frac{-2\alpha_1}{\kappa^2} h_i^b {\cal A}_i^b.
\end{equation}
From here one can derive the tensor-current correlator:
\begin{equation}
\vev{T^{<ij>}\, J_a^k} \simeq -\vev{\cal O}(\delta^{ik}\delta^j_a+\delta^i_a\delta^{jk}-\delta^{ij}\delta^k_a).
\end{equation}
When introduce this result in the Kubo formula \eqref{wetah} we find that $\widetilde{\eta}_H$ vanishes to this order.

\section{Conclusions}
\label{sec:conc}

Our original motivation for this work has been the question of  whether parity breaking due to locking
of internal and space symmetries in a superfluid phase leads generically to the relation between the Hall viscosity and the
angular momentum density
 $\frac{\eta_H}{\ell} = \frac{1}{2}$. For other sources of parity breaking it is known not to be true, as has been shown in several holographic models \cite{Jensen:2011xb,Wu:2013vya,Liu:2014gto}. The reason to suspect that this could be the case is the possibility that locking may imply the same origin
for the generation of both the Hall viscosity and the angular momentum density, thus linking their values.

In order to answer this question,
we studied the first order hydrodynamics of relativistic non-Abelian superfluid in $2+1$ dimensions, where
we locked the $SU(2)$ internal symmetry with the space symmetry.
Note as a side remark,  that the Goldstone bosons and the corresponding superfluid velocities pattern in a non-Abelian
superfluid depends on the pattern of symmetry breaking, and the study of the general case is worth pursuing  in the future.

The parity breaking due
to the locking generated parity-odd terms in the constitutive relations, which we analysed in detail.
In particular we studied the relation between the Hall viscosity and angular momentum density,
which turned out to be generically of order one, but not necessarily one half.
The holographic $p$-wave model studied in \cite{Son:2013xra} falls within our class of locked superfluid hydrodynamics,
and we showed that our general results are compatible with it. As part of our analysis we have derived a Kubo formula for the angular momentum density. We observe that it receives a contribution proportional to the Hall conductivity \eqref{c3}. This suggests that a similar formula exists for Abelian fluids, explaining the appearance of non-zero angular momentum density in holographic models with a Chern-Simons term for the dual gauge field \cite{Jensen:2011xb,Liu:2012zm,Wu:2013vya,Liu:2014gto}.

Finally, we demonstrated how locking corrects parity-even transport such as shear and bulk viscosities, and
also found that there are qualitatively different extensions of transports and in particular Hall viscosity to non-zero velocities.

\section*{Acknowledgements}

This work is supported in part by the Israeli Science Foundation Center
of Excellence, the US-Israel Binational Science Foundation, and by the I-CORE program of Planning and Budgeting Committee and the Israel Science Foundation (grant number 1937/12).

\appendix

\section{Modification of the entropy current}\label{modifent}

The canonical entropy current is
\begin{equation}
J^\mu_{\rm can} = s u^\mu -\frac{\pi^{\mu\nu}}{T}u_\nu -\nu_A^\mu \frac{\mu_A}{T}.
\end{equation}
The entropy current can have additional terms
\begin{equation}
J^\mu_s = J^\mu_{\rm can} + S^\mu.
\end{equation}
Where $S^\mu$ should be such that
\begin{equation}
\nabla_\mu J^\mu_s\geq 0.
\end{equation}
In principle having $S^\mu$ allows more independent first order terms than those allowed by the canonical entropy current alone, that for us were $\sigma^{\mu\nu}, \theta, v_A^\mu$ and $s_A$.

We make the following simplification. Both parity and time-reversal invariance are not broken explicitly, so there are no epsilon tensors appearing in the first order terms and Onsager's relations should be satisfied. This implies that there are no cancellations among first order terms in the divergence of the entropy current. Therefore,
\begin{equation}
\nabla_\mu J^\mu_s = \left(\text{canonical quadratic terms} \right)+\left(\text{new quadratic terms} \right)+\left(\text{cross terms} \right).
\end{equation}
The new quadratic terms can only come from the divergence of $S^\mu$.

In our case we expand
\begin{equation}
S^\mu = u^\mu \Sigma_u + N^\mu \Sigma_N + V^\mu,
\end{equation}
where $u_\mu V^\mu=N_\mu V^\mu=0$. For purely Abelian configurations, the analysis of \cite{Bhattacharya:2011tra} verified that there are no new independent terms when parity and time reversal invariance are unbroken and $S^\mu$ could be set to zero. This implies that possible new terms should be proportional to the non-Abelian superfluid velocity $\hat{\zeta}^\mu_A$ (terms depending on the chemical potential and external fields get an index but are otherwise the same). To the order we are working this means that the new quadratic terms must be quadratic in $\hat{\zeta}^\mu_A$, and it should be possible to write them as the square of new terms linear in the non-Abelian components of the superfluid velocity. If there are no such terms then the canonical entropy current will not be modified.

In the locked phase we set the Abelian velocity $N^\mu=0$ and the sources to zero, so we will check whether, in this subclass of configurations, there can be additional terms in the entropy current. The first derivative terms we can have are
\begin{equation}
\sigma_{\mu\nu},  \ \ \omega_{\mu\nu}=P_{\mu}^{\ \alpha}P_{\nu}^{\ \beta} (\partial_\alpha u_\beta - \partial_\beta u_\alpha),  \ \
a_\mu=u^\alpha \partial_\alpha u_\mu, \ \ \theta,  \ \ \partial_\mu\left(\frac{\mu_A}{T} \right), \ \ \partial_\mu T, \ \ \partial_\mu \hat{\zeta}^\nu_A.
\end{equation}
Then, at $O(\hat{\zeta}^2)$ in the scalar sector $\Sigma_u$ we have 6 terms
\begin{equation}
\begin{aligned}
&S_{i=1,\dots,6}=\hat{\zeta}^2 \times \left[\theta,  \ \ u^\alpha\partial_\alpha T, \ \  u^\alpha \partial_\alpha\left(\frac{\mu_A}{T} \right)\mu_A \right];  \ \ \zeta^{\mu\nu}\sigma_{\mu\nu},  \ \ \tilde{\zeta}^{\mu\nu}\omega_{\mu\nu}, \ \  u^\mu\partial_\mu\hat{\zeta}^2.
\end{aligned}
\end{equation}
In the vector sector $V^\mu$ we have 13 terms
\begin{equation}
\begin{aligned}
V_{i=1,\dots,13}^\mu &=\left[ \hat{\zeta}^2P^{\mu\alpha},  \ \ \zeta^{\mu\alpha}, \ \  \tilde{\zeta}^{\mu\alpha}\right]\times \left[ a_\alpha, \ \  \partial_\alpha T, \ \  \partial_\alpha\left(\frac{\mu_A}{T} \right)\mu_A\right]; \ \  P^{\mu\alpha}\partial_\alpha\hat{\zeta}^2,\\
&\quad \hat{\zeta}^\alpha_A P^{\mu\nu}\times \left[\partial_\alpha \hat{\zeta}_{\nu\,A}, \ \ \partial_\nu \hat{\zeta}_{\alpha\,A} \right]; \ \  \hat{\zeta}^\mu_A \partial_\alpha \hat{\zeta}^\alpha_A.
\end{aligned}
\end{equation}
Note that not all the terms are necessarily independent but they may be related by the ideal order equations of motion. 

We write the $O(\hat{\zeta}^2)$ contributions to the entropy as
\begin{equation}
\Sigma_u^{(2)} = \sum_i s_i(T,\mu_A) S_i, \ \ V^{(2)\,\mu} = \sum_i v_i(T,\mu_A) V_i^\mu,
\end{equation}
and
\begin{equation}
\begin{aligned}
\partial_\mu S^{(2)\,\mu} &= \sum_i \frac{\partial s_i}{\partial T} u^\mu\partial_\mu T S_i +\frac{\partial s_i}{\partial \mu_A/T} u^\mu\partial_\mu \left(\frac{\mu_A}{T} \right) S_i+s_i \theta S_i + s_i u^\mu\partial_\mu S_i\\
&+\sum_i \frac{\partial v_i}{\partial T} V_i^\mu\partial_\mu T  +\frac{\partial v_i}{\partial \mu_A/T} V_i^\mu\partial_\mu \left(\frac{\mu_A}{T} \right) + v_i \partial_\mu V_i^\mu.
\end{aligned}
\end{equation}
From this expression we see that the possible quadratic terms are, from the scalar terms
\begin{equation}
\hat\zeta^2 \left(u^\alpha\partial_\alpha T\right)^2, \ \ \hat\zeta^2 \left(u^\alpha\partial_\alpha \left(\frac{\mu_A}{T} \right)\mu_A\right)^2,\ \ \hat\zeta^2\theta^2, \ \ \left(u^\alpha\partial_\alpha \hat\zeta^\mu_A\right)^2.
\end{equation}
The last term comes from $u^\mu\partial_\mu S_6$.
From the vector terms we have
\begin{equation}
\hat\zeta^2\left(P^{\mu\alpha}\partial_\alpha T\right)^2,\ \hat\zeta^2 \left(P^{\mu\alpha}\partial_\alpha \left(\frac{\mu_A}{T} \right)\mu_A\right)^2 ,\ \ \left(\partial_\alpha\hat\zeta_A^\alpha\right)^2, \ \ \left(\partial_\alpha\hat \zeta_A^\beta\right)^2.
\end{equation}
The last two terms originate from the $\partial_\mu V_i^\mu$ terms. All these terms can be written as the square of the following first-order terms
\begin{equation}
\begin{split}
&\hat\zeta^\mu_A \times \left[ u^\alpha\partial_\alpha T, \ u^\alpha\partial_\alpha \left(\frac{\mu_B}{T} \right)\mu_B ,\ \theta\right],\ u^\alpha\partial_\alpha \hat\zeta^\mu_A\ \partial_\alpha\hat\zeta^\alpha_A,\\
&\hat\zeta^\mu_A \times \left[ P^{\nu\alpha}\partial_\alpha T, \ P^{\nu\alpha}\partial_\alpha \left(\frac{\mu_B}{T} \right)\mu_B\right], \ \partial_\mu \hat\zeta^\nu_A, \\
&\tilde\zeta^\mu_A \times \left[ u^\alpha\partial_\alpha T, \ u^\alpha\partial_\alpha \left(\frac{\mu_B}{T} \right)\mu_B ,\ \theta\right],\\
&\tilde\zeta^\mu_A \times \left[ P^{\nu\alpha}\partial_\alpha T, \ P^{\nu\alpha}\partial_\alpha \left(\frac{\mu_B}{T} \right)\mu_B\right].
\end{split}
\end{equation}
The terms in second and fourth lines 
are tensors with a global index $\tau^{\mu\nu}_A$, so they cannot appear in any of the dissipative contributions to the energy-momentum tensor or the current. The last term in the second line could appear in $H_A$, while the remaining terms could appear in $\nu_A^\mu$ (in fact the terms proportional to $\theta$ are already there, so they do not introduce anything new).

The only terms that in principle could affect to our discussion of parity breaking in the locked phases are then the first two terms in the third line, that are proportional to $\tilde\zeta^\mu_A$. However, their only effect is to add additional scalar contributions to the current, even after locking. Therefore, they do not affect to the Hall viscosity or the angular momentum. In \S\S~\ref{4a} the only effect is to add new terms in the scalar part below \eqref{scalar1} and in \eqref{scalar2}, only in the dissipative terms of the current (not in the energy-momentum tensor). For the calculations involving the angular momentum density, in \S\S~\ref{4b} the only non-vanishing terms are proportional to the shear viscosity, so the new terms are absent. In \S\S~\ref{4c} we consider static configurations with vanishing normal component of the velocity, so it is clear that the new terms also vanish. For the analysis of the Hall viscosity in \S\S~\ref{4d} we study the energy-momentum tensor, where the possible new terms do not enter.

\section{Dissipative terms}
\label{sec:diss}

We collect in this appendix the dissipative terms that can appear to $O(\epsilon^2)$, in a general non-Abelian superfluid and in the different locked phases.

\subsection{General non-Abelian superfluid}
\label{sec:diss1}

Using basic building blocks allowed by the entropy equation \eqref{bbs}
we can construct the following terms\footnote{Note that $\epsilon^{ABC}\epsilon^{A'B'C'}=\delta^{AA'}\delta^{BB'}\delta^{CC'}- \delta^{AB'}\delta^{BA'}\delta^{CC'} +\ \text{cyclic permutations}$.}
\begin{itemize}
\item Tensor $\tau^{\mu\nu}$
\begin{equation}
\begin{split}
& O(1) \ \ \sigma^{\mu\nu},\\
& O(\epsilon) \ \ v^\mu_A\hzeta^\nu_A, \ v^\mu_A\tzeta^\nu_A,\\
& O(\epsilon^2) \ \  \sigma^\mu_\alpha\zeta^{\alpha\nu}, \  \sigma^{\mu}_\nu\tzeta^{\alpha\nu}.
\end{split}
\end{equation}
\item Mixed tensor $\tau^\mu_A$
\begin{equation}
\begin{split}
& O(1) \ \ v^\mu_B P_{BA}, \ v^\mu_B C_{BA}\\
& O(\epsilon) \ \ \sigma^\mu_\alpha \hzeta^\alpha_A,\ \sigma^\mu_\alpha\tzeta^\alpha_A\\
& O(\epsilon^2) \ \  v_{\alpha\, B}\zeta^{\mu\alpha}P_{BA},\ v_{\alpha\, b}\zeta^{\mu\alpha}C_{BA},\ v_{\alpha B}\tzeta^{\mu\alpha}P_{BA}, v_{\alpha\,B}\hzeta^\mu_B\hzeta^\alpha_A ,\  v_{\alpha\,B}\hzeta^\mu_B\tzeta^\alpha_A,\  v_{\alpha\,B}\tzeta^\mu_B\hzeta^\alpha_A,\ v^\mu_B\zeta_{BA}.
\end{split}
\end{equation}
\item Vector $V^\mu$,
\begin{equation}
\begin{split}
& O(1) \ \ v^\mu_A\mu_A\\
& O(\epsilon) \ \ \sigma^\mu_\alpha N^\alpha,\  s_A\hzeta^\mu_A,\  s_A\tzeta^\mu_A\\
& O(\epsilon^2) \ \  v_{\alpha\, A}\zeta^{\mu\alpha}\mu_A,\  v_{\alpha A}\tzeta^{\mu\alpha}\mu_A,\  v_{\alpha\, A}N^\alpha\hzeta^\mu_A, \  v_{\alpha\, A}N^\alpha\tzeta^\mu_A.
\end{split}
\end{equation}
\item Internal vector $V_{A}$
\begin{equation}
\begin{split}
& O(1) \ \ s_B P_{AB}, \ s_B C_{BA},\\
& O(\epsilon) \ \ v_{\alpha B}N^\alpha P_{AB},\ v_{\alpha B}N^\alpha C_{AB},\ v_{\alpha B}\hzeta^\alpha_A \mu_B,\  v_{\alpha B}\tzeta^\alpha_A\mu_B, \\
& O(\epsilon^2) \ \   \sigma_{\alpha\beta}N^\alpha\hzeta^\beta_A, \  \sigma_{\alpha\beta}N^\alpha\tzeta^\beta_A,\ s_B\zeta_{AB}.
\end{split}
\end{equation}
\item Scalar $\Sigma$,
\begin{equation}
\begin{split}
& O(1) \ \ \theta,\ s_A\mu_A,\\
& O(\epsilon) \ \ v_{\alpha A}N^\alpha \mu_A,\   v_{\alpha A}\hzeta^\alpha_A,\  v_{\alpha A}\tzeta^\alpha_A, \\
& O(\epsilon^2) \ \  \sigma_{\alpha\beta}N^\alpha N^\beta, \sigma_{\alpha\beta}\zeta^{\alpha\beta}.
\end{split}
\end{equation}
\end{itemize}

\subsection{Locking in the lab frame}
\label{sec:diss2}

The terms that in principle survive after locking are
\begin{itemize}
\item Tensor $\tau^{\mu\nu}$
\begin{equation}
\begin{split}
& O(1) \ \ \sigma^{\mu\nu},\\
& O(\epsilon) \ \ \text{none},\\
& O(\epsilon^2) \ \  \sigma^\mu_\alpha\zeta^{\alpha\nu}, \  \sigma^{\mu}_\nu\tzeta^{\alpha\nu}.
\end{split}
\end{equation}
\item Mixed tensor $\tau^\mu_A$
\begin{equation}
\begin{split}
& O(1) \ \ \text{none}\\
& O(\epsilon) \ \ \sigma^\mu_\alpha \hzeta^\alpha_A,\ \sigma^\mu_\alpha\tzeta^\alpha_A\\
& O(\epsilon^2) \ \  \text{none}.
\end{split}
\end{equation}
\item Vector $V^\mu$,
\begin{equation}
\begin{split}
& O(1) \ \ v^\mu_A\mu_A\\
& O(\epsilon) \ \  s_A\hzeta^\mu_A,\  s_A\tzeta^\mu_A\\
& O(\epsilon^2) \ \  v_{\alpha\, A}\zeta^{\mu\alpha}\mu_A,\  v_{\alpha A}\tzeta^{\mu\alpha}\mu_A.
\end{split}
\end{equation}
\item Internal vector $V_{A}$
\begin{equation}
\begin{split}
& O(1) \ \ s_B P_{AB}, \ s_B C_{BA},\\
& O(\epsilon) \ \ v_{\alpha B}\hzeta^\alpha_A \mu_B,\  v_{\alpha B}\tzeta^\alpha_A\mu_B, \\
& O(\epsilon^2) \ \ \ s_B\zeta_{AB}.
\end{split}
\end{equation}
\item Scalar $\Sigma$,
\begin{equation}
\begin{split}
& O(1) \ \ \theta,\ s_A\mu_A,\\
& O(\epsilon) \ \  \text{none}, \\
& O(\epsilon^2) \ \  \sigma_{\alpha\beta}\zeta^{\alpha\beta}.
\end{split}
\end{equation}
\end{itemize}

\subsection{Locking in the rest frame}
\label{sec:diss3}

The terms that in principle survive after locking are
\begin{itemize}
\item Tensor $\tau^{\mu\nu}$
\begin{equation}
\begin{split}
& O(1) \ \ \sigma^{\mu\nu},\\
& O(\epsilon) \ \ v^\mu_A\hzeta^\nu_A, \ v^\mu_A\tzeta^\nu_A,\\
& O(\epsilon^2) \ \  \sigma^\mu_\alpha\zeta^{\alpha\nu}, \  \sigma^{\mu}_\nu\tzeta^{\alpha\nu}.
\end{split}
\end{equation}
\item Mixed tensor $\tau^\mu_A$
\begin{equation}
\begin{split}
& O(1) \ \ v^\mu_B P_{BA}, \ v^\mu_B C_{BA}\\
& O(\epsilon) \ \ \sigma^\mu_\alpha \hzeta^\alpha_A,\ \sigma^\mu_\alpha\tzeta^\alpha_A\\
& O(\epsilon^2) \ \  v_{\alpha\, B}\zeta^{\mu\alpha}P_{BA},\ v_{\alpha\, B}\zeta^{\mu\alpha}C_{BA},\ v_{\alpha B}\tzeta^{\mu\alpha}P_{BA}, v_{\alpha\,B}\hzeta^\mu_B\hzeta^\alpha_A ,\  v_{\alpha\,B}\hzeta^\mu_B\tzeta^\alpha_A,\  v_{\alpha\,B}\tzeta^\mu_B\hzeta^\alpha_A,\ v^\mu_B\zeta_{BA}.
\end{split}
\end{equation}
\item Vector $V^\mu$,
\begin{equation}
\begin{split}
& O(1) \ \ v^\mu_A\mu_A,\\
& O(\epsilon) \ \  s_A\hzeta^\mu_A,\  s_A\tzeta^\mu_A,\\
& O(\epsilon^2) \ \  v_{\alpha\, A}\zeta^{\mu\alpha}\mu_A,\  v_{\alpha A}\tzeta^{\mu\alpha}\mu_A.
\end{split}
\end{equation}
\item Internal vector $V_{A}$
\begin{equation}
\begin{split}
& O(1) \ \ s_B P_{AB}, \ s_B C_{BA},\\
& O(\epsilon) \ \ v_{\alpha B}\hzeta^\alpha_A \mu_B,\  v_{\alpha B}\tzeta^\alpha_A\mu_B, \\
& O(\epsilon^2) \ \ \ s_B\zeta_{AB}.
\end{split}
\end{equation}
\item Scalar $\Sigma$,
\begin{equation}
\begin{split}
& O(1) \ \ \theta,\ s_A\mu_A,\\
& O(\epsilon) \ \  v_{\alpha A}\hzeta^\alpha_A,\  v_{\alpha A}\tzeta^\alpha_A, \\
& O(\epsilon^2) \ \  \sigma_{\alpha\beta}\zeta^{\alpha\beta}.
\end{split}
\end{equation}
\end{itemize}

\section{Equations and action in the holographic $p$-wave model}
\label{sec:eqs}
In this appendix we collect some of the results of \cite{Son:2013xra}, in particular the equations of motion of the background and fluctuations and the form of the on-shell action.
The background metric and gauge field are charged black hole solutions of the form
\begin{equation}
\begin{split}
&ds^2=-F(z) dt^2+\frac{dz^2}{F(z)}+r^2(z)(dx^2+dy^2),\\
&A^3_{0}=\phi(z), \ \ A^a_i=A(z)\delta^a_{i}.
\end{split}
\end{equation}
As $z\to \infty$ the metric approaches asymptotically $AdS_4$.

\subsection{Background equations of motion}

We will denote derivatives with respect to $z$ with primes:
\begin{equation}
\begin{split}
& 2r r'' +(A')^2+\lambda^2\frac{\phi^2 A^2}{F^2}=0,\\
& F''-2\frac{F}{r^2}(r')^2-(\phi')^2-\lambda^2\frac{A^4}{r^4}=0,\\
& (r^2\phi')'-2\lambda^2\frac{\phi A^2}{F}=0,\\
& (FA')'+\lambda^2\left(\frac{\phi^2 A}{F}-\frac{A^3}{r^2} \right)=0,\\
&F''+4\frac{Fr''}{r}+2\frac{F (r')^2}{r^2}+4\frac{r' F'}{r}=48.
\end{split}
\end{equation}
Linearized equation for the background $A$:
\begin{equation}
(F {A^{(1)}}')'+\lambda^2\frac{\phi^2}{F}A^{(1)}=0.
\end{equation}
Asymptotic expansion
\begin{equation}
A^{(1)}(z)=\alpha_0^{(1)}+\frac{\alpha_1^{(1)}}{z}+\cdots.
\end{equation}

\subsection{On-shell action}

Fluctuations around the background solution are denoted as $\delta g_{\mu\nu}=h_{\mu\nu}$ and $\delta A_{A\,\mu}=a_{A\,\mu}$. Indices are raised and lowered with the background metric $\bar{g}_{\mu\nu}$. The trace of the fluctuation is denoted as $h=\bar{g}^{\mu\nu}h_{\mu\nu}$. Covariant derivatives $\bar{\nabla}$ are taken with respect to the background metric. The bulk contribution to the on-shell action is
\begin{equation}
\begin{split}
S_{\text{on-shell}} &=\lim_{z\to \infty}\frac{1}{4\kappa^2}\int d^3 x r^2\left[h\Gamma^{z}_{\nu\alpha} h^{\alpha\nu}-h^{\rho\sigma}\Gamma^z_{\rho\alpha} h^\alpha_\sigma+\frac{3}{2}h^{\rho\sigma}\bar{\nabla}^zh_{\rho\sigma}-\frac{1}{2}h\bar{\nabla}^z h\right.\\
&\left.-a_{A\,\nu}\left( F_A^{z\nu}h+F_{A\rho}^{\ \ z}h^{\nu\rho}+f_A^{z\nu}\right)\right].
\end{split}
\end{equation}
To this action one should add the boundary Gibbons-Hawking and counterterm actions:
\begin{equation}
\begin{split}
S_{GH} &= \lim_{z\to\infty} \int d^3x\, \sqrt{F}r^2\left((K+\bar{\nabla}_z)\left(\frac{1}{2}h^2-h^{\mu\nu} h_{\mu\nu} \right) \right),\\
S_{ct} &= -\lim_{z\to\infty}\frac{1}{\kappa^2}\int d^3x\, \sqrt{F}r^2\left(\frac{1}{2}h^2-h^{\mu\nu} h_{\mu\nu} \right).
\end{split}
\end{equation}
Where $K_{\mu\nu}$ is the extrinsic curvature of the background metric.

\subsection{Vector fluctuations}

We will restrict to time-independent fluctuations $\delta g_{\mu\nu}=h_{\mu\nu}(z)$, $\delta A_{A\,\mu}= a_{A\,\mu}(z)$. At zero momentum modes can be grouped in parity even and parity odd:
\begin{equation}
\begin{split}
&(h_{tx}+h_{ty})/2=r^2 h_t^e, \ \ (h_{tx}-h_{ty})/2=r^2 h_t^o,\\
&(a_{1\,t}+a_{2\,t})/2=a_t^e, \ \ (a_{1\,t}-a_{2\,t})/2= a_t^o,\\
&(a_{3\,x}+a_{3\,y})/2=a_3^e, \ \ (a_{3\,x}-a_{3\,y})/2=a_3^o.
\end{split}
\end{equation}
The equations for fluctuations $h_t^i,a_t^i, a_3^i$ are
\begin{equation}\label{veceqs}
\begin{split}
& (r^4 {h_t^i}')'+r^2\phi' {a_3^i}'=S_t^{h\,i},\\
& r^2\phi' {h_t^i}'+(F{a_3^i}')'=S_3^i,\\
& \left(\frac{{a_t^i}}{\phi}\right)'=S_t^{a\,i}.
\end{split}
\end{equation}
Where
\begin{equation}\label{veceqs2}
\begin{split}
S_t^{h\,i} &=\left(r^2(A')^2+\lambda^2\frac{A^4}{F} \right)h_t^i-\lambda^2\frac{\phi A^2}{F}a_3^i-r^2 A' {a_t^i}'+\lambda^2\frac{A^3}{F}a_t^i,\\
S_3^i &= -\lambda^2 \frac{\phi A^2}{F}h_t^i+\lambda^2 \frac{A^2}{r^2}a_3^i+\lambda^2\frac{\phi A}{F}a_t^i,\\
S_t^{a\,i} &=\left(\frac{A}{\phi}\right)'h_t^i+\frac{F}{r^2\phi^2}\left(A' a_3^i-A{a_3^i}' \right).
\end{split}
\end{equation}

\subsection{Tensor fluctuations}
We will consider time-dependent fluctuations $\delta g_{\mu\nu}=e^{-i\omega t}h_{\mu\nu}(z)$, $\delta A_{a\,\mu}=  e^{-i\omega t}a_{a\,\mu}(z)$. At zero momentum the modes can be grouped in parity-even and parity-odd:
\begin{equation}
\begin{split}
&h_{xy}=r^2 h_e, \ \ (a_{1\,y}+a_{2\,x})/2=a_e,\\
&(h_{xx}-h_{yy})/2=r^2 h_o, \ \ (a_{1\,x}-a_{2\,y})/2= a_o.
\end{split}
\end{equation}
The equations for fluctuations $h_i$ and $a_i$ are
\begin{equation}\label{tenseq}
\begin{split}
& (r^2 F h_i')'=S_i^h,\\
& (F a_i')+\lambda^2\frac{\phi^2}{F}a_i=S_i^a.
\end{split}
\end{equation}
Where
\begin{equation}\label{tenseq2}
\begin{split}
S_i^h &=-\left(\frac{\omega^2 r^2}{F}-F(A')^2+\lambda^2\frac{\phi^2A^2}{F} \right) h_i+2\left(-FA' a_i'+\lambda^2\frac{\phi^2 A}{F}a_i \right)+2\lambda i\omega\frac{\phi A}{F}\epsilon_{ij}a_j.\\
S_i^a &= FA'h_i'+\lambda^2\frac{A^3}{r^2}h_i-\left(\frac{\omega^2}{F}+\lambda^2\frac{A^2}{r^2}\right)a_i+\lambda i\omega\epsilon_{ij}\frac{\phi}{F}\left(A h_j-2a_j \right) .
\end{split}
\end{equation}
The renormalized quadratic on-shell action takes the simpler form:
\begin{equation}
S_{\text{on-shell}}=\lim_{z\to \infty}\frac{1}{2\kappa^2}\int d^3 x \sum_i \left[ -\frac{r^2 F}{2} h_i h_i'-Fa_i a_i'-\frac12\left((r^2 F)'-8r^2\sqrt{F} \right)h_i^2+FA' h_i a_i\right].
\end{equation}

\section{Calculation of Hall viscosity in the holographic $p$-wave}
\label{sec:corr}

We split the metric and gauge perturbations in leading order and corrections
\begin{equation}
h_i=H_i+\delta h_i, \ \ a_i=A_i+\delta a_i.
\end{equation}
The equations of motion for the leading order part are
\begin{equation}
\begin{split}
(r^2 F H_i')'+\frac{r^2\omega^2}{F}H_i=0,\\
(F A_i')'+\frac{\lambda^2\phi^2}{F}A_i+\frac{\omega^2}{F}A_i+2i\omega\frac{\lambda\phi}{F}\epsilon_{ij}A_j=0.
\end{split}
\end{equation}
We impose ingoing boundary conditions at the horizon $z\to z_H$
\begin{equation}
H_i \sim (z-z_H)^{-i\frac{\omega}{4\pi T}}, \ \ A_i \sim (z-z_H)^{-i\frac{\omega}{4\pi T}}.
\end{equation}
And at the boundary $z\to \infty$ the leading order terms are constant
\begin{equation}
H_i\simeq h_i^b, \ \ A_i \simeq a_i^b.
\end{equation}
The equations for the next order corrections are
\begin{equation}
\begin{split}
(r^2 F \delta h_i')'+\frac{r^2\omega^2}{F}\delta h_i=2\left(FA' A_i' +\frac{\lambda^2\phi^2}{F^2}A A_i\right)+2i\omega\frac{\lambda\phi}{F}A\epsilon_{ij} A_j,\\
(F \delta a_i')'+\frac{\lambda^2\phi^2}{F}\delta a_i+\frac{\omega^2}{F}\delta a_i+2i\omega\frac{\lambda\phi}{F}\epsilon_{ij}\delta a_j=F A' H_i'+i\omega \frac{\lambda\phi}{F}A\epsilon_{ij} H_j.
\end{split}
\end{equation}
In order to solve these equations we first write the corrections as
\begin{equation}
\delta h_i=H_i\sigma_i, \ \ \delta a_i=A_i\rho_i.
\end{equation}
Then,
\begin{equation}
\begin{split}
\sigma_i''+\left(2\frac{H_i'}{H_i} +\frac{(r^2 F)'}{r^2 F}\right)\sigma_i'=\frac{S_i^h(A_i)}{r^2 F H_i},\\
\rho_i''+\left(2\frac{A_i'}{A_i}+\frac{F'}{F} \right)\rho_i'=\frac{S_i^a(H_i)}{F A_i}.
\end{split}
\end{equation}
The `source' terms are
\begin{equation}
\begin{split}
&S_i^h(A_i)=2\left(FA' A_i' +\frac{\lambda^2\phi^2}{F^2}A A_i\right)+2i\omega\frac{\lambda\phi}{ F}A\epsilon_{ij} A_j,\\
&S_i^a(H_i)=F A' H_i'+i\omega \frac{\lambda\phi}{F}A\epsilon_{ij} H_j.
\end{split}
\end{equation}
The solutions are easily found
\begin{equation}
\begin{split}
&\sigma_i^{(1)}(z)=\int_{z_H}^z\frac{dz_1}{H_i^2 r^2 F}\int_{z_H}^{z_1}dz_2\,H_i S_i^h(A_i),\\
&\rho_i^{(1)}(z)=\int_{z_H}^z\frac{dz_1}{A_i^2 F}\int_{z_H}^{z_1}dz_2\,A_i S_i^a(H_i).
\end{split}
\end{equation}
The choice of integration limits is necessary in order to preserve the condition that the solutions are ingoing at the horizon.

This fixes the solution to first order, but we will also need the solution to second order, given by the equations
\begin{equation}
\begin{split}
{\sigma_i^{(2)}}''+\left(2\frac{H_i'}{H_i} +\frac{(r^2 F)'}{r^2 F}\right){\sigma_i^{(2)}}'=\frac{S_i^h(A_i,H_i,\rho_i^{(1)})}{r^2 F H_i},\\
{\rho_i^{(2)}}''+\left(2\frac{A_i'}{A_i}+\frac{F'}{F} \right){\rho_i^{(2)}}'=\frac{S_i^a(A_i,H_i,\sigma_i^{(1)})}{F A_i}.
\end{split}
\end{equation}
The `source' terms are
\begin{equation}
\begin{split}
S_i^h(A_i,H_i,\rho_i^{(1)}) &=2\left(FA' (A_i'\rho_i^{(1)}+A_i {\rho_i^{(1)}}') +\frac{\lambda^2\phi^2}{F^2}A A_i\rho_i^{(1)}\right)+2i\omega\frac{\lambda\phi}{ F}A\epsilon_{ij} A_j\rho_j^{(1)}\\
&+\left(F(A')^2-\frac{\lambda^2\phi^2}{F}A^2 \right)H_i,\\
S_i^a(A_i,H_i,\sigma_i^{(1)}) &=F A' (H_i'\sigma_i^{(1)}+H_i{\sigma_i^{(1)}}')+i\omega \frac{\lambda\phi}{F}A\epsilon_{ij} H_j \sigma_j^{(1)}-\frac{\lambda^2}{r^2}A^2 A_i.
\end{split}
\end{equation}
The solutions are then
\begin{equation}
\begin{split}
&\sigma_i^{(2)}(z)=\int_{z_H}^z\frac{dz_1}{H_i^2 r^2 F}\int_{z_H}^{z_1}dz_2\,H_i S_i^h(A_i,H_i,\rho_i^{(1)}),\\
&\rho_i^{(2)}(z)=\int_{z_H}^z\frac{dz_1}{A_i^2 F}\int_{z_H}^{z_1}dz_2\,A_iS_i^a(A_i,H_i,\sigma_i^{(1)}).
\end{split}
\end{equation}

\subsection{Boundary expansion}

The on-shell action for the tensor modes is
\begin{equation}
S=\lim_{z\to \infty} \frac{1}{2\kappa^2}\int d^3x\,\left[-\frac{r^2F}{2}h_ih_i'-F a_ia_i'-\frac{1}{2}((r^2F)'-8 r^2\sqrt{F} )h_i^2+FA' h_i a_i\right].
\end{equation}
We will use the expansion
\begin{equation}
F\simeq 4z^2, \ \ \phi \simeq \mu, \ \ \phi' \simeq -\frac{\phi_1}{z^2}, \ \ A\simeq \frac{\alpha_1}{z}, \ \ r^2\simeq 4 z^2, \ \ H_i\simeq h_i^b, \ \ H_i'\simeq-\frac{3 T_i}{z^4}, \ \ A_i\simeq a_i^b, \ \ A_i'\simeq -\frac{J_i}{z^2} .
\end{equation}
And
\begin{equation}
h_i\simeq {\cal H}_i+\frac{{\cal T}_i}{z^3}, \ \ a_i\simeq {\cal A}_i+\frac{{\cal J}_i}{z}.
\end{equation}
The on-shell action becomes
\begin{equation}
S=\int d^3x\,\left[\frac{28}{\kappa^2}{\cal H}_i{\cal T}_i,+\frac{2}{\kappa^2}{\cal A}_i{\cal J}_i-\frac{2\alpha_1}{\kappa^2}{\cal H}_i {\cal A}_i\right].
\end{equation}
The last term we already computed, we are interested in the term proportional to ${\cal T}_i$.

One can see that with the corrections
\begin{equation}
h_i=h_i^b(1+\sigma^{(1)}(\infty)+\sigma^{(2)}(\infty))+O(z^{-3}), \ \ a_i=a_i^b(1+\rho^{(1)}(\infty)+\rho^{(2)}(\infty))+O(z^{-1}).
\end{equation}
The sources in the dual field theory are then given by
\begin{equation}
{\cal H}_i=h_i^b(1+\sigma^{(1)}(\infty)+\sigma^{(2)}(\infty)), \ \ {\cal A}_i=a_i^b(1+\rho^{(1)}(\infty)+\rho^{(2)}(\infty)).
\end{equation}

The subleading terms can be computed as
\begin{equation}
\begin{split}
{\cal J}_i &=\lim_{z\to\infty}-z^2\partial_z a_i, \\
{\cal T}_i &=\lim_{z\to\infty}(-z^2\partial_z)^3 h_i=-\lim_{z\to\infty}(6 z^4 h_i'+6z^5 h_i''+z^6 h_i''').
\end{split}
\end{equation}

Note that the equation of $H_i$ does not have terms with $\omega\epsilon_{ij}$, therefore the contribution to the Hall viscosity vanishes to leading order. To next order, there is one contribution proportional to the epsilon tensor, but it involves the leading order gauge field solution $\sim \omega\epsilon_{ij}A_j$, so it will produce a term of the form $\sim \omega\epsilon_{ij}{\cal A}_i {\cal H}_j$. To second order there is a term $\sim \omega\epsilon_{ij} H_j$, that enters through  $\rho^{(1)}$. Therefore, the Hall viscosity is $\sim \vev{\cal O}^2$.

\end{document}